\documentclass[journal]{IEEEtran}
\usepackage{subfigure, amssymb, amsmath}
\usepackage{amsthm}
\usepackage{array, multirow}
\usepackage{graphics, graphicx}
\usepackage{epstopdf}
\usepackage{mathrsfs}
\usepackage{mathtools}
\usepackage{color}
\usepackage{url}
\usepackage{booktabs} % special rule in table
\usepackage{bm}
\usepackage[noadjust]{cite}
\usepackage{algorithm}
\usepackage{algpseudocode}
\usepackage[flushleft]{threeparttable}
\newcommand*{\Scale}[2][4]{\scalebox{#1}{$#2$}}%
\newtheorem{assumption}{Assumption}
\newtheorem{theorem}{Theorem}

\newtheorem{remark}{Remark}

\ifCLASSINFOpdf
\else
\fi
\begin{document}
\title{Stochastic Co-design of Storage and Control for Water Distribution Systems}

\author{Ye Wang, Erik Weyer, Chris Manzie, Angus R. Simpson, Lisa Blinco%Michael~Shell,~\IEEEmembership{Member,~IEEE,}
        %John~Doe,~\IEEEmembership{Fellow,~OSA,}
        %and~Jane~Doe,~\IEEEmembership{Life~Fellow,~IEEE}% <-this % stops a space
% \thanks{M. Shell was with the Department
% of Electrical and Computer Engineering, Georgia Institute of Technology, Atlanta,
% GA, 30332 USA e-mail: (see http://www.michaelshell.org/contact.html).}% <-this % stops a space
% \thanks{J. Doe and J. Doe are with Anonymous University.}% <-this % stops a space
% \thanks{Manuscript received April 19, 2005; revised August 26, 2015.}
\thanks{Y. Wang, E. Weyer and C. Manzie are with Department of Electrical and Electronic Engineering, The University of Melbourne, Parkville VIC 3010, Australia. {\tt\small  E-mail: \{ye.wang1, ewey, manziec\}@unimelb.edu.au}}
\thanks{A. R. Simpson is with School of Architecture and Civil Engineering, University of Adelaide, SA 5005, Australia. {\tt\small E-mail: angus.simpson@adelaide.edu.au}}
\thanks{L. Blinco is with South Australian Water Corporation, SA 5005, Australia. {\tt\small E-mail: Lisa.Blinco@sawater.com.au}}
}

\maketitle

\begin{abstract}
Water distribution systems (WDSs) are typically designed with a conservative estimate of the ability of a control system to utilize the available infrastructure. The controller is designed and tuned after a WDS has been laid out, a methodology that may introduce unnecessary conservativeness in both system design and control, adversely impacting operational efficiency and increasing economic costs. To address these limitations, we introduce a method to simultaneously design infrastructure and develop control parameters, the co-design problem, with the aim of improving the overall efficiency of the system. Nevertheless, the co-design of a WDS is a challenging task given the presence of stochastic variables (e.g. water demands and electricity prices). In this paper, we propose a tractable stochastic co-design method to design the best tank size and optimal control parameters for WDS, where the expected operating costs are established based on Markov chain theory. We also give a theoretical result showing that the average long-run operating cost converges to the expected operating cost with probability~1. Furthermore, this method is not only applicable to greenfield projects for the co-design of WDSs but can also be utilized to improve the operations of existing WDSs in brownfield projects. The effectiveness and applicability of the co-design method are validated through three illustrative examples and a real-world case study in South Australia. 
\end{abstract}


% \fi
\IEEEpeerreviewmaketitle

%%%%%%%%%%%%%%%%%%%%%%%%%%%%%%%%%%%%%%%%%%%%%%%%%%%%%%%
%%%%%%%%%%%%%%%%%%%%%%%%%%%%%%%%%%%%%%%%%%%%%%%%%%%%%%%
\section{Introduction}\label{section:introduction}

% Background and Motivation
\IEEEPARstart{R}{eliable} and continuous water supply is of vital importance for all activities in modern cities and rural communities. Water distribution systems (WDS) are critical infrastructure used to supply water from sources (e.g. reservoirs, rivers, groundwater or desalination plants) through pressurized pipes to customers. Given the physical distances spanned by water networks,  storage tanks are typically used to increase the robustness of the overall system to faults, disruptions (e.g., due to maintenance), and fluctuations in supply and demand. In Australia, energy consumption for water distribution has been predicted to be as high as 201 TWh by 2025 \cite{IEA}.

As in many other countries, the wholesale energy market in Australia is operated by a national agency, the Australian Energy Market Operator (AEMO). The wholesale market, which may be accessed by water authorities as well as energy retailers, is characterized by time-varying electricity spot prices. A challenge is consequently to design and utilize the available infrastructure to meet water demand while also minimizing the combination of infrastructure and operating costs in the presence of variable energy pricing. This is complicated by the wide range of time scales in the problem - the electricity prices vary in the order of minutes, yet the infrastructure is fixed for decades.

% Design of WDSs
Optimization techniques are often employed to balance infrastructure investment and operational savings. Evolutionary algorithms have been suggested to address water distribution system design \cite{savic1997genetic, marques2018many, batchabani2014optimal}. Other objectives, such as reducing greenhouse gas emissions, have also been considered in the design process in \cite{wu2013multiobjective}. However, most of these approaches do not consider an explicit control policy but an approximation of a potential operating cost.

% Control of WDSs
With a designed infrastructure, effective control operations can provide cost efficiency without compromising water supply \cite{cembrano2000optimal,wang2017non,puig2017real,Guo2023}. Over the past two decades, optimization-based techniques (e.g. model predictive control) have been widely investigated in academia, see e.g., \cite{zheng2015energy,wang2021minimizing,mala2017lost,sampathirao2017gpu,giuliani2017scalable,creaco2019real,salomons2020practical,oikonomou2020optimal}. These works consider the situation where the infrastructure already exists and the objective is to optimize its utilization in delivering water when and where required. The physical infrastructure is directly or indirectly reflected as a constraint(s) in the control problem, and so has an impact on the efficiency of the day-to-day operations. 

% Emphasizing the importance of the codesign problem
 Recently, the co-design of infrastructure and control in WDSs has gained attention \cite{pecci2017outer,garcia2019control}. However, it is challenging to implement such a co-design approach for WDSs under long-term uncertainties. In our previous study \cite{Wang-2022}, we presented preliminary results of a simplified co-design problem optimizing both the tank size and a simplified control policy under constant water demands. The approach utilized Markov chain theory \cite{meyn2012markov,shiryayev1996probability} to analyze total co-design cost under stochastic electricity prices \cite{prabhu1998stochastic}. The resulting optimization problem was tractable, leading to optimal co-design solutions. 

The main contribution of this paper is to propose a tractable stochastic co-design method for simultaneously optimizing the selection of the storage tank size and control parameters for WDSs. We consider an aggregated WDS that captures the main features of WDSs. Water demands and electricity prices are stochastic. To handle these stochastic characteristics, we use Markov chain theory \cite{bertsekas2008introduction,shiryayev1996probability} to analyze the evolution of the volume of water in the storage tank, which depends on both the size of the tank and the control policy. Furthermore, the control policy from the co-design solution can also be applied to existing WDS to improve operational performance. We provide three examples and a real case study in South Australia to illustrate and demonstrate the proposed method. 

The remainder of this paper is organized as follows: The co-design problem is described in Section \ref{section:problem statement}. A stochastic co-design method is proposed in Section \ref{section:codesign}. In Sectio \ref{section:examples}, three examples are provided to illustrate the proposed co-design optimization method. The results for a real case study are presented in Section \ref{section:case study} before conclusions are drawn in Section~\ref{section:conclusions}.

\paragraph*{Notation} 

{We use $\mathbf{E}[\cdot]$ and $\mathbf{Pr}\{\cdot\}$ to denote the mathematical expectation and probability, respectively.} For a stochastic variable $r$, the probability density function (PDF) is denoted by $f(r)$ and the cumulative distribution function (CDF) is denoted by $F(r)$. $r \sim \mathcal{N}(\mu, \sigma^2) $ means that $r$ is normally distributed with mean $\mu$ and variance $\sigma^2$. In this case, 
\begin{align*}
    f(r) &= \frac{1}{\sigma \sqrt{2\pi}} e^{-\frac{1}{2}\left (\frac{r-\mu}{\sigma}\right)^2},\\
    F(r) &= \int_{-\infty}^{r} f(t) dt = \frac{1}{2} \left [1+\mathrm{erf} \left (\frac{r-\mu}{\sigma \sqrt{2}}\right) \right],
\end{align*}
where $\mathrm{erf}(\cdot)$ is Gauss error function. Furthermore, we use $\mathbf{0}$ to denote a matrix of suitable dimension with all elements equal to zero. For two integers $a$ and $b$ ($b$ non-zero), $\mathrm{mod}(a,b)$ denotes the modulo operation that returns the remainder of a division of $a$ by $b$. {$\mathbb{R}_{+}$ denotes the nonnegative real numbers while and $\mathbb{N}$ denotes the natural numbers.}

%%%%%%%%%%%%%%%%%%%%%%%%%%%%%%%%%%%%%%%%%%%%%%%%%%%%%%%
%%%%%%%%%%%%%%%%%%%%%%%%%%%%%%%%%%%%%%%%%%%%%%%%%%%%%%%
\section{Problem Description}\label{section:problem statement}

\begin{figure}[htpb]
    \centering
    \includegraphics[width=0.9\hsize]{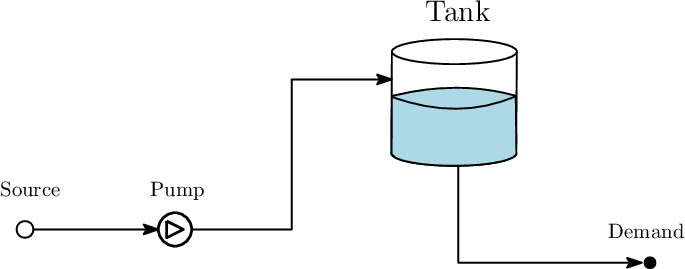}
    \caption{The topology of an aggregated water distribution system.}
    \label{fig:example}
\end{figure}

An aggregated WDS is composed of a water source, a pump, a water storage tank (of size $V$, to be designed) and a water demand sector, as shown in Fig. \ref{fig:example}. This aggregated model captures important features of realistic WDSs. In this paper, we use this aggregate system to investigate the co-design task.

The dynamics of the water volume in the tank can be described using volume balance as
\begin{align}\label{eq:tank dynamics}
    x_{k+1} = x_{k} + \Delta t \Big ( u_k \left( r_k, \alpha_k(x_k), x_k \right) - d_k \Big ),
\end{align}
where {$ x_k \in \mathbb{R}_{+} $} is the water volume in the tank as the system state, {$ u_k \in \mathbb{R}_{+} $} is the pumping flow to the tank as the control input, and {$d_k \in \mathbb{R}_{+} $} is the water demand as the exogenous input, at a time $k \in \mathbb{N}$. $\Delta t$ is the sampling interval and $k \in \mathbb{N}$ is a discrete time.

The water demand $d_k$ is uncertain. In practice, periodicities in the water demand can usually be observed, for instance, a daily or weekly pattern. The control policy is essentially to pump if the electricity price is below a price threshold~$\alpha_k(x_k)$. This threshold is allowed to be time-varying, reflecting the periodicities in water demands and electricity prices. It also depends on the volume of water in the tank reflecting that we are more inclined to pump when the tank is nearly empty than when there is plenty of water in the tank. Hence, one of the main tasks considered in this paper is the design of the price threshold function denoted by $\alpha_k(\cdot)$. The control policy $u_k(r_k,\alpha_k(x_k),x_k)$ determining whether to pump or not then becomes a function of $r_k$, $\alpha_k$ and $x_k$. To comply with operational requirements, the control policy is modified when the tank is nearly empty or nearly full. To this end, let $\overline{x} \leq V $ denote a maximum water volume dictated by operational constraints while $\underline{x} > 0$ denotes a minimum water volume to be kept in reserve. The control policy is described as follows: 
\begin{enumerate}
    \item If the electricity price is equal to or lower than the price threshold and the water volume in the tank is below or equal to an upper limit $\overline{x}$, then pumping occurs;
    \item If the tank is close to empty, then pumping has to occur irrespective of electricity price to comply with a minimum water storage requirement;
    \item Otherwise no pumping occurs.
\end{enumerate}

Overall, this control policy is
\begin{equation}\label{eq:pumping flow}
    u_k \left( r_k, \alpha_k(x_k),x_k \right) = \begin{cases}
     q, & \text{if } r_k \leq \alpha_k(x_k) \text{ and } x_k \leq \overline{x}, \\
     q, & \text{if }  x_k \leq \underline{x}, \\
     0, & \text{otherwise},
    \end{cases}
\end{equation}
where $q$ denotes a constant flow provided by the pump. 

The aggregated WDS has both capital and operating costs. The objective for the co-design problem is to minimize these costs by simultaneously designing the tank size $V$ and the control policy while considering a long-term planning horizon $N>0$ (the number of discrete-time steps). The overall co-design cost is given by
\begin{align}\label{eq:overall co-design cost}
    c_t(V) + \sum_{k=0}^{N} \ell_k(r_k, \alpha_k(x_k),x_k), 
\end{align}
where {$c_t(V)$ is the capital cost of the storage tank. The operating cost $\ell_k(r_k, \alpha_k(x_k),x_k)$ at time $k$ depends on the electricity price, price threshold and current state.}

Since $r_k$ and $d_k$ are stochastic variables, the cost in \eqref{eq:overall co-design cost} is a random variable. As $N$ is large, the long-term expected average cost is minimized, and the following co-design problem for the system is considered:
\begin{subequations}\label{problem:general co-design}
	\begin{align}
	& \underset{\substack{V, \; {\alpha_k(\cdot)}, \\k= 0,\ldots, N}}{\mathrm{minimize}} \;\; \frac{1}{N} c_t(V) + \mathbf{E} \left [ \frac{1}{N} \sum_{k=0}^{N} \ell_k(r_k, \alpha_k(x_k),x_k)  \right ], \label{eq:general codesign cost function}\\
	%%%%%%%%%%%%%%%%
	\intertext{subject to \eqref{eq:tank dynamics}, \eqref{eq:pumping flow} and}
	& x_0 = \tilde{x},\label{eq:initial tank volume constraint}\\
	& 0 \leq x_k \leq V, \; k = 0, \ldots, N, \label{eq:state constraint}
	\end{align}
\end{subequations}
where $\tilde{x}$ denotes an initial water volume in the tank, which can also be treated as a stochastic variable with a certain distribution. The expectation is with respect to $r_k$, $d_k$ and~$\tilde{x}$.

The co-design optimization problem in~\eqref{problem:general co-design} presents considerable challenges in terms of its tractability due to multiple factors. One significant issue is the potentially long planning horizon $N$, which often corresponds to the lifespan of infrastructure. Planning horizons of 20, 50, and even up to 100 years are common. This long-term outlook causes problems for conventional solution methodologies like the Monte Carlo tree search. Such approaches tend to become impractically cumbersome over these extended timescales. 

%%%%%%%%%%%%%%%%%%%%%%%%%%%%%%%%%%%%%%%%%%%%%%%%%%%%%%%
%%%%%%%%%%%%%%%%%%%%%%%%%%%%%%%%%%%%%%%%%%%%%%%%%%%%%%%
\section{Stochastic Co-design Optimizations based on Finite-state Markov Chain}\label{section:codesign}

In this section, we formulate the stochastic optimization problem based on Markov chain theory to find the storage tank size $V$ and {the price threshold function $\alpha_k(\cdot)$}. The reformulated optimization problem presented in this section is more tractable than \eqref{problem:general co-design} due to the approximations in the derivation of a finite-state Markov chain.

\subsection{Quantized Demands, Pumping Flows and Volumes}\label{subsection:quantization}

As introduced in Section \ref{section:problem statement}, water demands $d_k$ are time-varying and their distribution is periodic with period $T$. {From the dynamics \eqref{eq:tank dynamics}, it can be seen that the system state (water volume in the tank) depends on water demands. The quantization of water demands and pumping flows allows us to represent the dynamics as a finite-state Markov chain.}

\begin{assumption}\label{assump:demands}
    The stochastic water demand $d_k$, $\forall k \in \mathbb{N}$ can take one of $m$ finite values, which can be represented as multiples of some positive scalar $d$, i.e.
    \begin{equation}\label{eq:dk def}
        d_k = \tau_k d \;\;\text{with probability of } a^{\tau_k}_k, 
    \end{equation}
    where $ \tau_k = 0,1,\ldots,m-1$. Moreover, the water demands at different time instances are independent of each other. It holds $a^{\tau}_{k} = a^{\tau}_{k+T}$ for any $\tau=0,\ldots, m-1$. Furthermore, the probabilities $a^{\tau}_{k}$, $\tau = 0,\ldots, m-1$ satisfy for every $k \in \mathbb{N}$,
    \begin{subequations}\label{eq:demand probabilities}
        \begin{align}
            &a^{\tau}_{k} \geq 0, \allowdisplaybreaks \\
            &\sum_{\tau=0}^{m-1} a^{\tau}_{k} = 1.
        \end{align}
    \end{subequations}
\end{assumption}

A close approximation of demands can be achieved if $d$ is chosen to be small, but it would lead to more states in the Markov chain to be introduced in the next subsection. We make the following assumption on the pumping flow based on $d$.

\begin{assumption}\label{assump:flow}
    {It is assumed that there is always water available for pumping.} The pumping flow~$q$ is a multiple of~$d$, i.e.
    \begin{equation}\label{eq:q def}
        q = \zeta d,
    \end{equation}
    where $\zeta>0$ is an integer.
\end{assumption}

From Assumption \ref{assump:flow}, it follows that the actual pumping flow in \eqref{eq:pumping flow} can be reformulated as
\begin{align}
    u_k\left( r_k, \alpha_k(x_k),x_k \right) = \zeta_k d,
\end{align}
where $\zeta_k = \zeta $ if pumping occurs, otherwise $\zeta_k = 0$.

Let $ \Delta x = d \Delta t $. From Assumptions \ref{assump:demands}-\ref{assump:flow}, it follows that at each time step there exists an integer $ \gamma $ (positive, negative or zero) such that
\begin{align}\label{eq:delta x}
    x_{k+1} - x_{k} = \gamma d \Delta t = \gamma \Delta x.
\end{align}

In addition, the following two assumptions are made for establishing the Markov chain.

\begin{assumption}\label{assump:volumes}
    The total volume $V$, the volume limits $\overline{x}$, $\underline{x}$ and the initial water volume $\tilde{x}$ are multiples of $\Delta x$, i.e. $V = n \Delta x$, $ \underline{x}= n_p\Delta x $, $ \overline{x} = n_s \Delta x $ and $\tilde{x} = n_0 \Delta x$, where $n$, $n_p$, $n_s$ and $n_0$ are integers.
\end{assumption}

\begin{assumption}\label{assump:electricity price Gaussian}
    The electricity prices $r_k$, $\forall k \in \mathbb{N}$ are stochastic and conditioned on time $k$. Moreover, they do admit probability densities $f_k(r_k)$ that are periodically varying in time $k$ with a period of $T$. Furthermore, given time $k$, the electricity price $r_k$ is independent of all other electricity prices, as well as the water demand and the initial water volume in the tank.
\end{assumption}

%%%%%%%%%%%%%%%%%%%%%%%%%%%%%%%%%%%%%%%%%%%%%%%%%%%%%%%
\subsection{Finite-state Markov Chain of Volume Evolution}\label{section: MC}

From Assumptions \ref{assump:demands}-\ref{assump:volumes}, water volume in the tank is a multiple of $\Delta x$. Moreover, due to the independence assumptions (Assumptions \ref{assump:demands} and \ref{assump:electricity price Gaussian}), the dynamics~\eqref{eq:tank dynamics} and the control policy \eqref{eq:pumping flow}, {the system can be represented as a finite state Markov chain and the states in the Markov chain are represented as follows:
\begin{align}\label{eq:Markov chain states}
    z_k : = [z_{1_k},z_{2_k}]^\top = [i_k, \kappa_k]^\top,
\end{align}
where $ i_k = \frac{x_k}{\Delta_x}$ represents water volume in the tank at time $k$ and $\kappa_k = \mathrm{mod}(k,T)$ is the time step within the period of $T$.}

Also from Assumption \ref{assump:volumes}, the maximum value of $i$ is $ n = \frac{V}{\Delta x}  $ corresponding to a full tank while the minimum value is~0 corresponding to an empty tank. 

\begin{figure}[ht]
    \centering
    \includegraphics[width=0.8\hsize]{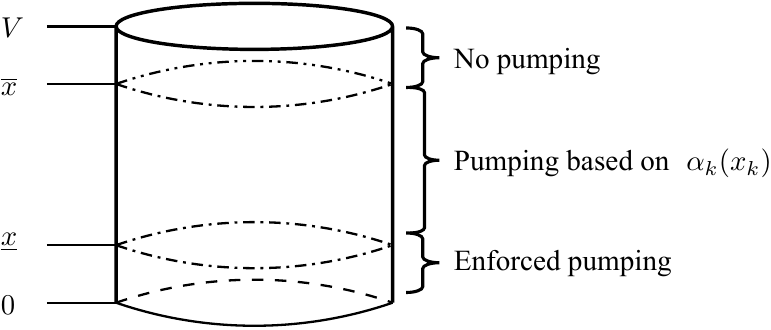}
    \caption{Labeled water volumes in the tank and associated control actions.}
    \label{fig:water volumes in tank}
\end{figure}

Fig. \ref{fig:water volumes in tank} shows water volumes with associated control actions using the control policy \eqref{eq:pumping flow}. $\overline{x}$ is chosen such that the tank will not overflow if there is pumping and the water demand is zero, i.e. $n_{s}+\zeta \leq n$. Similarly, we also consider that the tank is large enough so that it does not empty if there is no pumping when the water volume in the tank is above $\bar{x}$ and the water demand is maximum, i.e. $n_{p} - m  + 1 \geq 0$. The total number of states in the Markov chain is $ \bar{n} = (n+1) T$. 

\begin{figure}[t]
    \centering
    \includegraphics[width=\hsize]{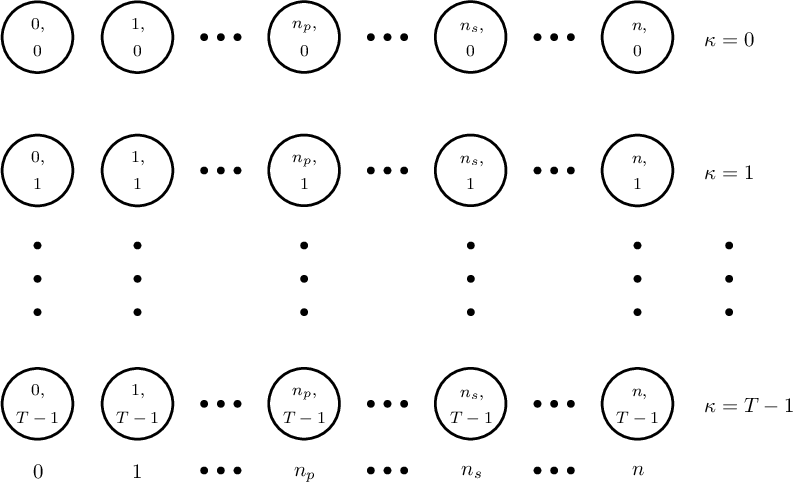}
    \caption{{The states in the Markov chain.}}
    \label{fig:MC full}
\end{figure}

Fig. \ref{fig:MC full} shows that the states $(i,\kappa)$ for $i = 0,\ldots, n$ and $\kappa = 0,\ldots, T-1$ in the Markov chain. The row represents a time between 0 and $T-1$ and the column represents water volumes in the tank between $0$ and $V$.

In the following, we discuss the transition probabilities for states in the Markov chain. The transition probability from state $(i,\kappa)$ to state $(j, \mathrm{mod}(\kappa + 1,T))$ at time $\kappa+1$ is denoted as $p_{\kappa}^{i,j}$. 

\subsubsection{States with no pumping}

According to the control policy introduced in \eqref{eq:pumping flow}, no pumping occurs when the water volume in the tank is above $\overline{x} = n_s\Delta x$.

\begin{figure}[ht]
    \centering
    \includegraphics[width=0.6\hsize]{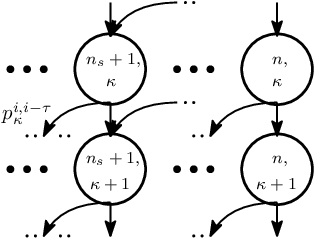}
    \caption{{States and transition probabilities with no pumping.}}
    \label{fig:MC Case 4}
\end{figure}

As shown in Fig.~\ref{fig:MC Case 4}, for states $(i,\kappa)$, $ i = n_s+1, \ldots, n $ for every $\kappa = 0,\ldots, T-1$, the transition probabilities from state $(i,\kappa)$ to state $(i-\tau,\mathrm{mod} (\kappa+1,T)) $ is
\begin{align}
     p^{i,i-\tau}_{\kappa} &= a^{\tau}_{\kappa}, \; \tau = 0,1,\ldots, m-1,\label{eq:prob4}
\end{align}
where $a^{\tau}_{\kappa}$ is given in \eqref{eq:dk def}.

\subsubsection{States with pumping based on $\alpha_{\kappa}(i\Delta x)$}

For states~$(i,\kappa)$, $ i = n_p+1, \ldots, n_s $ for every $\kappa = 0,\ldots, T-1$, pumping occurs if $ r_k \leq \alpha_{\kappa}(i \Delta x)$ with $ \kappa = \mathrm{mod}(k,T) $.

\begin{figure}[ht]
    \centering
    \includegraphics[width=0.9\hsize]{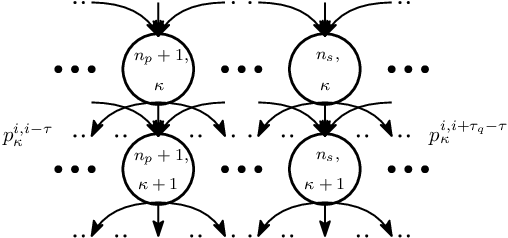}
    \caption{{States and transition probabilities with pumping based on $\alpha_{\kappa}(i \Delta x)$.}}
    \label{fig:MC Case 3}
\end{figure}

When $r_k \leq \alpha_{\kappa}(i\Delta x)$, the next state $(j,\mathrm{mod} (\kappa+1,T))$ in the Markov chain when the water demand is $\tau d$, is obtained from 
\begin{align}
    j = i + \zeta - \tau.
\end{align}

Therefore, the transition probability from state $(i, \kappa) $ to state $(i + \zeta - \tau,\mathrm{mod} (\kappa+1,T))$ is
\begin{align}\label{eq:transition probability-alpha-pumping}
    p^{i,i + \zeta - \tau}_{\kappa} = F_{\kappa}(\alpha_{\kappa}(i \Delta x)) a^{\tau}_{\kappa}, \; \tau=0,1,\ldots, m-1,
\end{align}
where $F_{\kappa}(\cdot)$ is the cumulative distribution function for electricity prices.

When $r_k > \alpha_{\kappa}(i\Delta x)$, no pumping occurs. The transition probability from state $(i,\kappa)$ to state $(i-\tau,\mathrm{mod} (\kappa+1,T))$ is
\begin{align}\label{eq:transition probability-alpha-nopumping}
    p^{i,i-\tau}_{\kappa} = \Big( 1-F_{\kappa}(\alpha_{\kappa}(i \Delta x)) \Big) a^{\tau}_{\kappa},\; \tau = 0,1,\ldots, m-1.
\end{align}

\subsubsection{States with enforced pumping}

According to the control policy introduced in \eqref{eq:pumping flow}, enforced pumping is triggered when the water volume in the tank is at or below $ \underline{x} = n_p\Delta x $. 

\begin{figure}[ht]
    \centering
    \includegraphics[width=0.75\hsize]{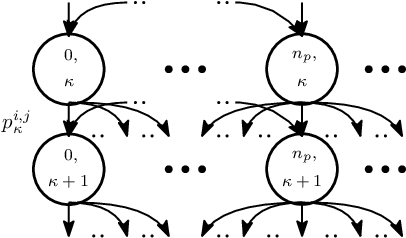}
    \caption{{States and transition probabilities with enforced pumping.}}
    \label{fig:MC Case 2}
\end{figure}

As shown in Fig.~\ref{fig:MC Case 2}, for states $(i,\kappa)$, $ i = 0, 1, \dots, n_p$ for every $\kappa = 0,\ldots, T-1$, due to that enforced pumping occurs, the next state in the Markov chain when the water demand is $\tau d$, can be found from
\begin{align}\label{eq:j compute pumping}
    j = \max \left(0, i+\zeta-\tau \right).
\end{align}

Therefore, the transition probability from state $(i,\kappa)$ to state $(j, \mathrm{mod} (\kappa+1,T))$ is
\begin{align}\label{eq:transition probability 2}
    p^{i,j}_{\kappa} = 
    \begin{cases}
        a^{i+\zeta-j}_{\kappa}, & \text{if } 0<j<i+\zeta-1, \\
        \displaystyle \sum_{\tau = i+\zeta}^{m} a^{\tau}_{\kappa}, & \text{if } j = 0.
    \end{cases}
\end{align}

%%%%%%%%%%%%%%%%%%%%%%%%%%
By stacking the states in Fig.~\ref{fig:MC full} row by row into a state vector, the following transition probability matrix is obtained
\begin{align}\label{eq:transition matrix}
    \mathbf{P} = \begin{bmatrix}
        \mathbf{0} & P_0 & \mathbf{0} & \ldots & \mathbf{0} & \mathbf{0}\\
        \mathbf{0} & \mathbf{0} & P_1 & \ldots & \mathbf{0} & \mathbf{0}\\
        \vdots & \vdots & \vdots & \ddots &\vdots & \vdots \\
        \mathbf{0} & \mathbf{0} & \mathbf{0} & \ldots & P_{T-2} & \mathbf{0}\\
        P_{T-1} & \mathbf{0} & \mathbf{0} & \ldots & \mathbf{0} & \mathbf{0}
    \end{bmatrix},
\end{align}
where for $\kappa = 0,\ldots, T-1$, 
\begin{align}
    P_{\kappa} = \begin{bmatrix}
        p^{0,0}_{\kappa} & \ldots & p^{0,n}_{\kappa}\\
        \vdots & \ddots & \vdots \\
        p^{n, 0}_{\kappa} & \ldots & p^{n,n}_{\kappa}
    \end{bmatrix},
\end{align}
and the transition probabilities $p^{i,j}_{\kappa} $, $ i,j = 0,1,\ldots, n $, $\kappa = 0,\ldots, T-1$ are given in \eqref{eq:prob4}, \eqref{eq:transition probability-alpha-pumping}, \eqref{eq:transition probability-alpha-nopumping} and \eqref{eq:transition probability 2}.

We make the following assumption on $\mathbf{P}$.

\begin{assumption}\label{assump:irr}
    The transition matrix $\mathbf{P}$ in \eqref{eq:transition matrix} is irreducible\footnote{Let $\tilde{p}_{(s)}^{i,j}$ be the probability of transiting from state $i$ to state $j$ in $s$ steps. A transition matrix $\mathbf{P}$ is irreducible if for any two states $i$ and $j$, there exist $s$ and $s'$ such that $\tilde{p}_{(s)}^{i,j} > 0$ and $\tilde{p}_{(s')}^{j,i} > 0$.}.
\end{assumption}

This assumption means that any state in the Markov chain can be reached from any other state in a finite number of steps with positive probability. This is a very mild assumption, which essentially means that any water volume $i\Delta x$ can be reached from any starting volume $j \Delta x$.

{Let
\begin{align*}
    \bm{\alpha} = [ \alpha_{\kappa} (i \Delta x), i = n_p+1, \ldots, n_s, \kappa = 0,\ldots, T-1 ]^\top
\end{align*} be price thresholds, where each price threshold depends on the state $(i,\kappa)$ in the Markov chain.}

{Let
\begin{align*}
    \bm{\pi}{(\bm{\alpha},V)} = [\pi^{i}_{\kappa}, i = 0,\ldots, n, \kappa = 0,\ldots, T-1]
\end{align*}
be the stationary probabilities of the states in the Markov chain, which} can be obtained from the balance and normalization equations:
\begin{subequations}
    \begin{align}
        \bm{\pi}{(\bm{\alpha},V)} &= \mathbf{P}^\top \bm{\pi}{(\bm{\alpha},V)},\label{eq:balance eq}\\
        1 &= \mathbf{1}_{\bar{n}} ^{\top} \bm{\pi}{(\bm{\alpha},V)},\label{eq:normalization eq}
    \end{align}
\end{subequations}
where $\mathbf{1}_{\bar{n}}$ is a vector with all $\bar{n} = (n+1)T$ elements equal to~1.

For an irreducible $\mathbf{P}$, the stationary probabilities are unique. 

{We next show the convergence with probability 1 of the time-averaged operating cost
\begin{align}\label{eq:total operating cost}
    W(N) := \frac{1}{N}\sum_{k=0}^{N-1}\ell_k(r_k, \alpha_k(x_k),x_k),
\end{align}
where $\ell_k(\cdot)$ is the operating cost function at time $k$ in \eqref{eq:overall co-design cost}.}

A key point is that conditioned on being in a particular state, the costs incurred in that state form a sequence of independent and identically distributed random variables. Further details and explanations are given in the proof below. For a given state, $i$ and $\kappa$ are fixed and the price $r_k$ is the only random variable. Note also that when in state $(i,\kappa)$, $k$ is limited to the values such that $\kappa = \mathrm{mod}(k,T) $. For these values of $k$, $r_k$ are i.i.d random variables with probability density $f_\kappa$ as given in Assumption \ref{assump:electricity price Gaussian}.

Let us denote the expected value (with respect to the density $f_\kappa$) of the cost while in this state by
\begin{align}
    \bar{\ell}_\kappa(\alpha_{\kappa}(i \Delta x), i \Delta x) = \mathbf{E}\Big[\ell_{\kappa} (\cdot, \alpha_{\kappa}(i \Delta x), i \Delta x)\Big].
\end{align}

\begin{theorem}\label{theorem:long-run property}
    For a given state $(i,\kappa)$, assume that the cost function $\ell_\kappa(\cdot, \alpha_{\kappa}(i \Delta x), i \Delta x)$ has finite second-order moment.
    
    Let
    \begin{align}\label{eq:ell bar}
        \bar{\ell} := \sum_{\kappa = 0}^{T-1} \sum_{i = 0}^{n} \pi_{\kappa}^{i} \bar{\ell}_\kappa(\alpha_{\kappa}(i \Delta x), i \Delta x).
    \end{align}
    Then, using the control policy \eqref{eq:pumping flow} under Assumptions \ref{assump:demands}-\ref{assump:irr}, it holds that for any distribution of initial state,
    \begin{align}
        \mathbf{Pr} \left \lbrace \lim_{N \rightarrow \infty} W(N) = \bar{\ell} \right \rbrace = 1.
    \end{align}
\end{theorem}

\begin{proof}
    The idea of the proof is to apply a strong law of large numbers to establish convergence with probability 1 to an expected value conditioned on being in a given state. This will be combined with a result from the Markov chain theory saying that for an irreducible Markov chain the relative number of visits to a state will converge with probability 1 to the stationary probability of the state. The limit in \eqref{eq:ell bar} is hence obtained by multiplying the expected value with the stationary probability of that state and then summing over all states.

    The state of the Markov chain at time $k$ is $ z_k = [z_{1_k},z_{2_k}]^\top = \left[\frac{x_k}{\Delta_x},\mathrm{mod}(k,T) \right ]^\top $. It follows that $
        z_{k+1} = \left[z_{1_k} + \zeta_k - \tau_k,\mathrm{mod}(z_{2_k} + 1,T) \right ]^\top
   $.
   
    From \eqref{eq:total operating cost}, we have
    \begin{align*}
        W(N) = \frac{1}{N} \sum_{k = 0}^{N-1}  \ell_{z_{2_k}}(r_k, \alpha_{z_{2_k}}(z_{1_k} \Delta x),z_{1_k} \Delta x).
    \end{align*}

    We next focus on a particular state $(i,\kappa)$ and introduce the indication functions%Let $\mathbf{I}_{i,\kappa} (z_k) $ be the indicator function:
    \begin{align*}
        \mathbf{I}_{i,\kappa} (z_k) = \begin{cases}
            1, & \text{if } z_{1_k} = i \text{ and } z_{2_k} = \kappa,\\
            0, & \text{otherwise. }
        \end{cases}
    \end{align*}

    Multiplication with the indicator functions allows us to sum over all states without changing the value of $W(N)$, and we obtain that%Then, the average operating cost is equivalent to
    \begin{align*}
        W(N) &= \frac{1}{N} \sum_{k = 0}^{N-1} \sum_{\kappa = 0}^{T-1} \sum_{i=0}^{n} \Scale[0.93]{\mathbf{I}_{i,\kappa} (z_k) \ell_{z_{2_k}} (r_k, \alpha_{z_{2_k}}(z_{1_k} \Delta x),z_{1_k} \Delta x)}\\
        &= \frac{1}{N} \sum_{k = 0}^{N-1} \sum_{\kappa = 0}^{T-1} \sum_{i=0}^{n} \mathbf{I}_{i,\kappa} (z_k) \ell_{\kappa} (r_k, \alpha_{\kappa}(i \Delta x), i \Delta x).
    \end{align*}
    
    Notice that the multiplication with the indicator function has allowed us to replace $\mathbf{I}_{i,\kappa} (z_k) \ell_{z_{2_k}} (r_k, \alpha_{z_{2_k}}(z_{1_k} \Delta x),z_{1_k} \Delta x)$ with $\mathbf{I}_{i,\kappa} (z_k) \ell_{\kappa} (r_k, \alpha_{\kappa}(i \Delta x), i \Delta x)$ in the above equation.

    Next, for the fixed state $(i,\kappa)$, we focus on the time average
    \begin{align}\label{eq:proof theorem 1 time average}
        \frac{1}{N} \sum_{k = 0}^{N-1} \mathbf{I}_{i,\kappa} (z_k) \ell_{\kappa} (r_k, \alpha_{\kappa}(i \Delta x), i \Delta x).
    \end{align}
    
    Let $s_{i,\kappa} = \sum_{k=0}^{N-1} \mathbf{I}_{i,\kappa} (z_k)$ denote the number of times the state $(i,\kappa)$ has been visited between time 0 and $N-1$. Let $\tilde{k}_{i,\kappa} (\rho)$ be the time index when the state~$(i,\kappa)$ is visited for the $\rho$-th time. Then, \eqref{eq:proof theorem 1 time average} can be reformulated as
    \begin{align*}
        &\frac{1}{N}\sum_{k = 0}^{N-1} \mathbf{I}_{i,\kappa} (z_k) \ell_{\kappa} (r_k, \alpha_{\kappa}(i \Delta x), i \Delta x)\\
        = \; & \frac{s_{i,\kappa}}{N} \cdot \frac{1}{s_{i,\kappa}} \sum_{\rho=0}^{s_{i,\kappa}}\ell_{\kappa} (r_{\tilde{k}_{i,\kappa}(\rho)}, \alpha_{\kappa}(i \Delta x), i \Delta x),
    \end{align*}
    where $\frac{s_{i,\kappa}}{N}$ is the relative number of visits to state $ (i,\kappa) $. From Assumption \ref{assump:irr} and \cite[Theorem 1.9.7]{suhov2008probability}, for any initial state, $\frac{s_{i,\kappa}}{N} \rightarrow \pi_{\kappa}^{i} $ with probability 1 as $N \rightarrow \infty $. It follows that $s_{i,\kappa} \rightarrow \infty$ as $N \rightarrow \infty$ when $\pi_{\kappa}^i > 0$.
    
    The term $\displaystyle\frac{1}{s_{i,\kappa}} \sum_{\rho=0}^{s_{i,\kappa}}\ell_{\kappa} (r_{\tilde{k}_{i,\kappa}(\rho)}, \alpha_{\kappa}(i \Delta x), i \Delta x)$ is the time average cost incurred when visiting state $(i,\kappa)$. Under Assumptions \ref{assump:demands} and \ref{assump:electricity price Gaussian} and due to that $z_k$ only depends on $z_{k-1}$, $r_{k-1}$ and $d_{k-1}$, $r_k$ is independent of $z_k, z_{k-1}, \ldots, z_0$. Therefore, $\ell_{\kappa} (r_{\tilde{k}_{i,\kappa}(\rho)}, \alpha_{\kappa}(i \Delta x), i \Delta x)$, $\rho=1, 2,\ldots$ is a sequence of independent random variables, {and $r_{\tilde{k}_{i,\kappa}(\rho)}$ is distributed according to the probability density $ f_\kappa $ in Assumption \ref{assump:electricity price Gaussian}.}
    
    As the cost functions $\ell_{\kappa} (r_{\tilde{k}_{i,\kappa}(\rho)}, \alpha_{\kappa}(i \Delta x), i \Delta x)$, $\rho = 1,2,\ldots$ have finite second-order moment, from Kolmogorov's strong law of large numbers \cite[Theorem 4.3.2]{shiryayev1996probability}, it follows that
    \begin{align*}
         \frac{1}{s_{i,\kappa}} \sum_{\rho=0}^{s_{i,\kappa}} \ell_{\kappa} (r_{\tilde{k}_{i,\kappa}(\rho)}, &\alpha_{\kappa}(i \Delta x), i \Delta x) \\ \rightarrow \; & \bar{\ell}_\kappa(\alpha_{\kappa}(i \Delta x), i \Delta x), 
    \end{align*}
    with probability 1 as $s_{i,\kappa} \rightarrow \infty$.

    Combining this with the convergence of $\frac{s_{i,\kappa}}{N}$ to $\pi_{\kappa}^{i} $ with probability 1 as $N \rightarrow \infty $, we obtain for fixed $i$ and $\kappa$,
    \begin{align*}
        \frac{1}{N} \sum_{k = 0}^{N-1} \mathbf{I}_{i,\kappa} (z_k) \ell_{\kappa} (r_k, &\alpha_{\kappa}(i \Delta x), i \Delta x) \\
         \rightarrow \; & \pi_{\kappa}^{i} \bar{\ell}_\kappa(\alpha_{\kappa}(i \Delta x), i \Delta x),
    \end{align*}
    with probability 1.

    Finally, by summing over all the states, we obtain that
    \begin{align*}
        W(N) \rightarrow \sum_{\kappa = 0}^{T-1} \sum_{i=0}^{n} \pi_{\kappa}^{i} \bar{\ell}_\kappa(\alpha_{\kappa}(i \Delta x), i \Delta x),
    \end{align*}
    with probability 1 as $N \rightarrow \infty$.
\end{proof}

Theorem \ref{theorem:long-run property} is useful for investigating the operating cost over a long horizon in a computationally efficient manner. The cost $\bar{\ell}$ in \eqref{eq:ell bar} will be used in the next subsection to approximate the operating cost. Note that the result in Theorem~\ref{theorem:long-run property} does not depend on the initial state of the Markov chain. Therefore, the constraint \eqref{eq:initial tank volume constraint} is omitted in the formulation in the following subsection. Furthermore, one iteration of the Markov chain takes one discrete-time sampling interval, such that there is a one-to-one correspondence between the real-time and the iteration index in the Markov chain.

%%%%%%%%%%%%%%%%%%%%%%%%%%%%%%%%%%%%%%%%%%%%%%%%%%%%%%%
\subsection{Stochastic Co-design Formulation}\label{section:codesign formulation}

Considering the Markov chain described above, we next reformulate the co-design cost function. The decision variables in the co-design optimization are the tank size $V$ and the state-dependent price thresholds in $ \bm{\alpha} = [ \alpha_{\kappa} (i \Delta x), i = n_p+1, \ldots, n_s, \kappa = 0,\ldots, T-1 ]^\top $. 

\subsubsection{Capital Cost}
The capital cost of a storage tank, $c_t(V)$, is a deterministic and monotonic function of volume.

\subsubsection{Operating Cost}

The operating cost includes two parts: the pumping energy cost and a penalty cost when the tank is empty or close to empty. The total pumping cost over the planning horizon of~$N$ time samples can be evaluated by the following two energy costs:
\begin{itemize}
    \item When the volume is $n_p \Delta x$ or lower, enforced pumping occurs based on the control policy in \eqref{eq:pumping flow}. In this case, the pump operates regardless of the price. The expected pumping energy cost is
    \begin{align}\label{eq:le}
        \bar{\ell}^{e}_{\kappa}(\alpha_{\kappa}(i \Delta x), i \Delta x) = \varepsilon_p \mu_\kappa,
    \end{align}
    for $\kappa = 0,\ldots, T-1$ and $i = 0,\ldots, n_p$, otherwise $\bar{\ell}^{e}_{\kappa}(\alpha_{\kappa}(i \Delta x), i \Delta x) = 0$, where $ \varepsilon_p $ denotes the energy consumption for pumping in a sampling interval when the pump is operating. $\mu_{\kappa}$ is the expected value of the electricity price at time $\kappa = \mathrm{mod}(k,T) $.
    \item When the volume is between $(n_p+1) \Delta x$ and $n_s \Delta x$, pumping occurs only when the price is below a threshold. {There is no pumping cost if the pump is not operating. According to the control policy in \eqref{eq:pumping flow}, the pump is operating with probability $\bar{p}^{i}_{\kappa}$.} Therefore, the expected pumping energy cost is
    \begin{align}\label{eq:lr}
        \bar{\ell}^{r}_{\kappa}(\alpha_{\kappa}(i \Delta x), i \Delta x) =  \bar{p}^{i}_{\kappa} \varepsilon_p \bar{c}_{\kappa}(\alpha_{\kappa} (i \Delta x)),
    \end{align}
    for $\kappa = 0,\ldots, T-1$ and $i = n_p +1,\ldots, n_s$, otherwise $\bar{\ell}^{r}_{\kappa}(\alpha_{\kappa}(i \Delta x), i \Delta x) = 0$, where
    \begin{subequations}
        \begin{align}
            \bar{c}_{\kappa}(\alpha_{\kappa} (i \Delta x)) := & \; \int_{-\infty}^{\alpha_{\kappa} (i \Delta x)} r f_{\kappa}(r) dr,\\
            \bar{p}^{i}_{\kappa}  = & \;F_{\kappa}(\alpha_{\kappa} (i \Delta x)),
        \end{align}
    \end{subequations}
    and $\bar{c}_{\kappa}(\alpha_{\kappa} (i \Delta x))$ is the expected value of the electricity price given that it is less than $\alpha_{\kappa} (i \Delta x)$. $ \bar{p}^{i}_{\kappa} $ denotes the probability that the pump is operating.
\end{itemize}

\begin{remark}
    If the electricity prices are Gaussian random variables with mean $\mu$ and variance $\sigma^2$, then the expected value of the electricity price given that it is less than or equal to $\alpha_{\kappa} (i \Delta x)$ is {\cite{bertsekas2008introduction}}
    \begin{align*}
        \bar{c}(\alpha_{\kappa} (i \Delta x)) := & \;\mathbf{E} \left[ r \mid r \leq \alpha_{\kappa} (i \Delta x)\right]\\
        = & \;\mu - \sigma^2\frac{f(\alpha_{\kappa} (i \Delta x))}{F(\alpha_{\kappa} (i \Delta x))}.
    \end{align*}
\end{remark}

Furthermore, low tank water volumes, especially an empty tank, should be avoided as they compromise the ability of the WDS to meet demand and are associated with a penalty. To incorporate this, a penalty cost when the volume of water is below $n_r \Delta x$ ($n_r\geq 0$) is applied. The penalty cost is given by
\begin{align}\label{eq:lp}
    \bar{\ell}^{p}_{\kappa}((\alpha_{\kappa}(i \Delta x), i \Delta x) =  w,
\end{align}
for $\kappa = 0,\ldots, T-1$ and $i = 0,\ldots, n_r$, otherwise $\bar{\ell}^{p}_{\kappa}((\alpha_{\kappa}(i \Delta x), i \Delta x) = 0$, where $w$ is the penalty. It applies every time instant the volume in the tank is at or below $ n_r \Delta x$.

%%%%%%%%%%%%%%%%%%%%%%%%%%%%%%%%%%%%%%%%%%%%%%%%%%%%%%%

From Theorem \ref{theorem:long-run property}, the operating cost
\begin{align*}
    \frac{1}{N} \sum_{k=0}^{N} \ell_k(r_k, \alpha_k(x_k),x_k)
\end{align*}
converges with probability 1 to
\begin{align}\label{eq:ell bar alpha V}
    \bar{\ell} (\bm{\alpha},V) = \sum_{\kappa = 0}^{T-1} \sum_{i = 0}^{n} \pi_{\kappa}^{i} \bar{\ell}_\kappa(\alpha_{\kappa}(i \Delta x), i \Delta x),
\end{align}
{where we have explicitly indicated the dependence on $\bm{\alpha}$ and $V$, and}
\begin{align*}
    \bar{\ell}_\kappa(\alpha_{\kappa}(i \Delta x), i \Delta x) = &\bar{\ell}^{e}_{\kappa}(\alpha_{\kappa}(i \Delta x), i \Delta x) + \bar{\ell}^{r}_{\kappa}(\alpha_{\kappa}(i \Delta x), i \Delta x) \\
    &+ \bar{\ell}^{p}_{\kappa}((\alpha_{\kappa}(i \Delta x), i \Delta x),
\end{align*}
where $\bar{\ell}^{e}_{\kappa}(\cdot)$, $\bar{\ell}^{r}_{\kappa}(\cdot)$ and $\bar{\ell}^{p}_{\kappa}(\cdot)$ are given in \eqref{eq:le}, \eqref{eq:lr} and \eqref{eq:lp}.

{As Theorem \ref{theorem:long-run property} holds for any distribution for initial states, the constraint \eqref{eq:initial tank volume constraint} is automatically satisfied. The constraint~\eqref{eq:state constraint} also holds due to the assumptions made above.} Therefore, the co-design problem \eqref{problem:general co-design} can be reformulated as 
\begin{align}\label{problem:stochastic co-design}
    \underset{\bm{\alpha}, V} {\mathrm{minimize}}\;\; & \frac{1}{N} c_t(V)  + \bar{\ell} (\bm{\alpha},V), 
\end{align}
where $\bar{\ell} (\bm{\alpha},V)$ is given in \eqref{eq:ell bar alpha V} and $ \bm{\alpha} = [ \alpha_{\kappa} (i \Delta x), i = n_p+1, \ldots, n_s, \kappa = 0,\ldots, T-1 ] $. 

{Summation of \eqref{problem:stochastic co-design}  over $N$ samples leads to a total co-design cost calculated by:
\begin{equation}\label{eq:total co-design cost}
  J(\bm{\alpha}, V) :=  c_t(V)  + N\bar{\ell} (\bm{\alpha},V).
\end{equation}}

\begin{remark}
    {If an annual inflation, $\beta$, is considered, the price threshold $\bm{\alpha}$ must be rescaled in year $j$ according to $\bm{\alpha} \to (1+\beta)^j\bm{\alpha}$. Nonetheless, if the inflation rate and the discount rate, $\xi$, are the same, the cost in \eqref{eq:total co-design cost} represents the overall net present value of a project. However, if a different discount rate $\xi \not = \beta$, and $K$ samples per year are considered, the net present value can be updated with
    \begin{align}
        J_{NPV}(\bm{\alpha}, V) := c_t(V) + K\bar{\ell} (\bm{\alpha},V) \displaystyle\sum_{j = 1}^{N/K} \frac{ (1+\beta)^j}{(1+\xi)^j}.
    \end{align}}
\end{remark}

%%%%%%%%%%%%%%%%%%%%%%%%%%%%%%%%%%%%%%%%%%%%%%%%%%%%%%%
\subsection{Stochastic Co-design Algorithm}

From the co-design formulation in \eqref{problem:stochastic co-design}, it can be seen that the number of decision variables in the vector $ \bm{\alpha} $ depends on the tank size $V$. For a given tank size $V_{\eta}$ with $\eta = 1,\ldots,s$, the optimization problem \eqref{problem:stochastic co-design} is solved by a numerical optimization algorithm, {e.g., simultaneous perturbation stochastic approximation (SPSA), where numerical gradient approximation can be computed by two measurements of total cost functions~\cite{spall2005introduction}}. Then, the tank size $V$ and the corresponding $\alpha$ that minimizes the co-design cost are obtained. We summarize the above procedure in Algorithm \ref{algorithm:co-design}. 

\begin{algorithm}[thbp]
    \caption{Stochastic Co-design of Storage and Control}
    \label{algorithm:co-design} 
    \begin{algorithmic}[1]
        \State Given possible tank sizes $V_1,\ldots, V_s$;
        \For {$\eta=1:s$}
            \State Obtain $\bm{\alpha}_\eta := \underset{\bm{\alpha}} {\arg \min}\; \bar{\ell} (\bm{\alpha},V_\eta)$ and its optimal cost $\bar{\ell}(\bm{\alpha}_\eta, V_{\eta})$; 
                \State Compute the total co-design cost $J(\bm{\alpha}_{\eta}, V_{\eta}) $ by \eqref{eq:total co-design cost};
        \EndFor
            \State Obtain the minimum co-design cost $J^*$ over all the possible tank sizes by $J^* := \min(J_1,\ldots, J_{s})$;
            \State Extract the optimal price thresholds $\bm{\alpha}^*$ and tank size $V^*$ corresponding to $J^*$.
    \end{algorithmic}
\end{algorithm}

%%%%%%%%%%%%%%%%%%%%%%%%%%%%%%%%%%%%%%%%%%%%%%%%%%%%%%%%
%%%%%%%%%%%%%%%%%%%%%%%%%%%%%%%%%%%%%%%%%%%%%%%%%%%%%%%%
\section{Illustrative Examples}\label{section:examples}

In this section, we provide three examples illustrating the proposed co-design method. In these three examples, the electricity prices are Gaussian random variables $r \sim \mathcal{N}(20, 10)$. The period is $ T= 1$, i.e. the distributions of water demand and electricity price do not vary with time. The capital cost of the storage tank is $ c_v =  10,000 $ per unit volume. The planning horizon is chosen as $20$ years that corresponds to $N=175,200$ samples using a sampling time interval of $\Delta t = 1$ hour. In the first example, the price threshold is state-independent and hence constant. In the second and third examples, state-dependent price pumping thresholds are considered.

%%%%%%%%%%%%%%%%%%%%%%%%%%%%%%%%%%%%%%%%%%%%%%%%%%%%%%%
\subsection{State-independent Price Threshold and Constant Demand}

\begin{figure*}
    \centering
    \begin{equation}\label{eq:pi0 case 1}
        \pi^0 = \begin{cases}
        \frac{1}{2V}, & \text{if } \alpha = \mu,\\
        \frac{F(\alpha) \left(1-2F(\alpha) \right) \left(1-F(\alpha) \right)^{V-1}}{F(\alpha) (1-2F(\alpha))(1-F(\alpha))^{V-1} + F(\alpha)^{V} (1-2F(\alpha)) + F(\alpha) (1-F(\alpha))^{V-1} - F(\alpha)^V} ,& \text{otherwise}.
        \end{cases}
    \end{equation}
    \hrule
\end{figure*}

\begin{figure}[H]
	\centering
	\includegraphics[width=\hsize]{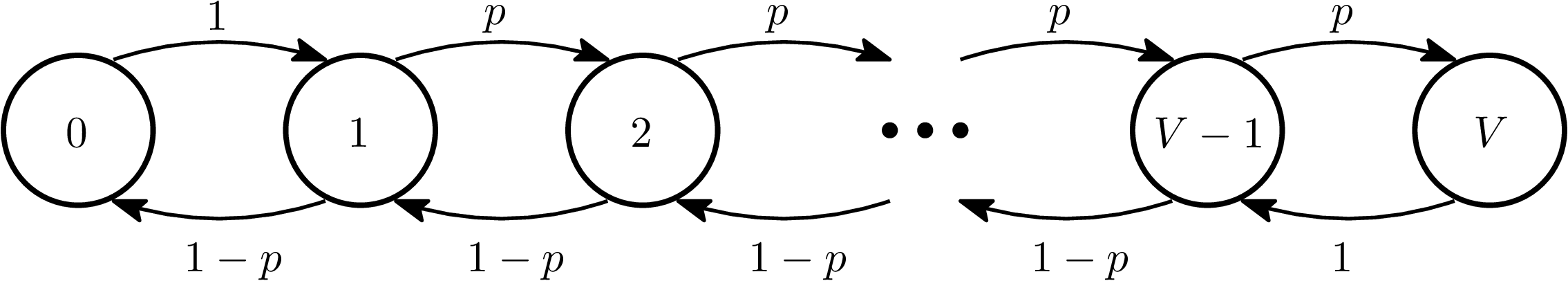}
	\caption{Markov chain for the example with state-independent price threshold and constant demand.}
	\label{fig:Case 1 markov chain}
\end{figure}

We first consider the case where the water demand is constant $d_k = d = 1$ volume unit per sampling interval. The pumping inflow is {$ q = \zeta = 2 $} volume unit per sampling interval when the pump is operating. The possible tank volume, $x$, ranges between 0 and a positive integer $V$. We consider the control policy described in \eqref{eq:pumping flow} with state-independent price threshold $\alpha$ and $\underline{x} = 0$ and $\overline{x} = V-1 $, {so $n_p = 0$ and $n_s = V-1$}. The corresponding Markov chain is shown in Fig.~\ref{fig:Case 1 markov chain}. The transition probabilities $p = F(\alpha)$ are the same since $\alpha$ is state-independent, where $F(\cdot)$ is the CDF of the Gaussian distribution.

The Markov chain in Fig. \ref{fig:Case 1 markov chain} is irreducible. The stationary probabilities depend on pumping probability~$p$. The stationary probabilities of the states, denoted by $\pi^i$, $i = 0,1,\ldots, V$, can be obtained from the following normalization and balance equations:
\begin{align*}
    &\sum_{i=0}^{V} \pi^{i} = 1,\allowdisplaybreaks\\
    &\pi^0 = (1-p) \pi^{1},\allowdisplaybreaks\\
    &\pi^{i} = p \pi^{i-1} + (1-p) \pi^{i+1}, \;\; i = 1,\ldots,V-2,\allowdisplaybreaks\\
    &\pi^{V-1} =  p \pi^{V-2} + \pi^V, \allowdisplaybreaks\\
    &\pi^V = p \pi^{V-1}.\allowdisplaybreaks
\end{align*}

For this example, we can explicitly derive an analytical expression for $\pi^0$ for the state $x=0$ parameterized by $\alpha$ and $V$. {After performing algebraic operation, the expression in \eqref{eq:pi0 case 1} is obtained and the stationary probabilities for the subsequent states $\pi^i$, $i = 1,\ldots, V$ can also be computed.} 

The operating cost can be divided into two cases: one due to enforced pumping with probability 1 from zero state $ i = 0 $; and pumping based on the price threshold $\alpha$ with probability $p = F(\alpha)$. Following the steps in Section \ref{section:codesign formulation} with $n_r = 0$, the expected operating cost is thus given by
\begin{align}\label{eq:operating cost}
   \bar{\ell}(\alpha,V) = & \pi^0(\alpha,V) (\varepsilon_p \mu+w) \nonumber\\
   & + \sum_{i=1}^{V-1} \pi^{i}(\alpha,V) p \varepsilon_p \bar{c}(\alpha), 
\end{align}
where $\bar{c} (\alpha) = \mu - \sigma^2\frac{f(\alpha)}{F(\alpha)}$ and $\varepsilon_p$ is the energy consumption in a sampling interval when the pump is operating. {Here $\varepsilon_p = 1$ and no penalty (i.e. $w= 0$) are used.}

Then, the stochastic co-design optimization problem can be formulated as follows:
\begin{align}\label{prob:stochastic codesign 1}
    (\alpha^*, V^*) & := 
    \underset{\alpha, V}{\arg \min}{\;\;\frac{1}{N} c_v V + \bar{\ell}(\alpha,V)},
\end{align}
where $\bar{\ell}(\alpha,V)$ is given in \eqref{eq:operating cost}.

\begin{figure}[thbp]
    \centering
    \includegraphics[width=\hsize]{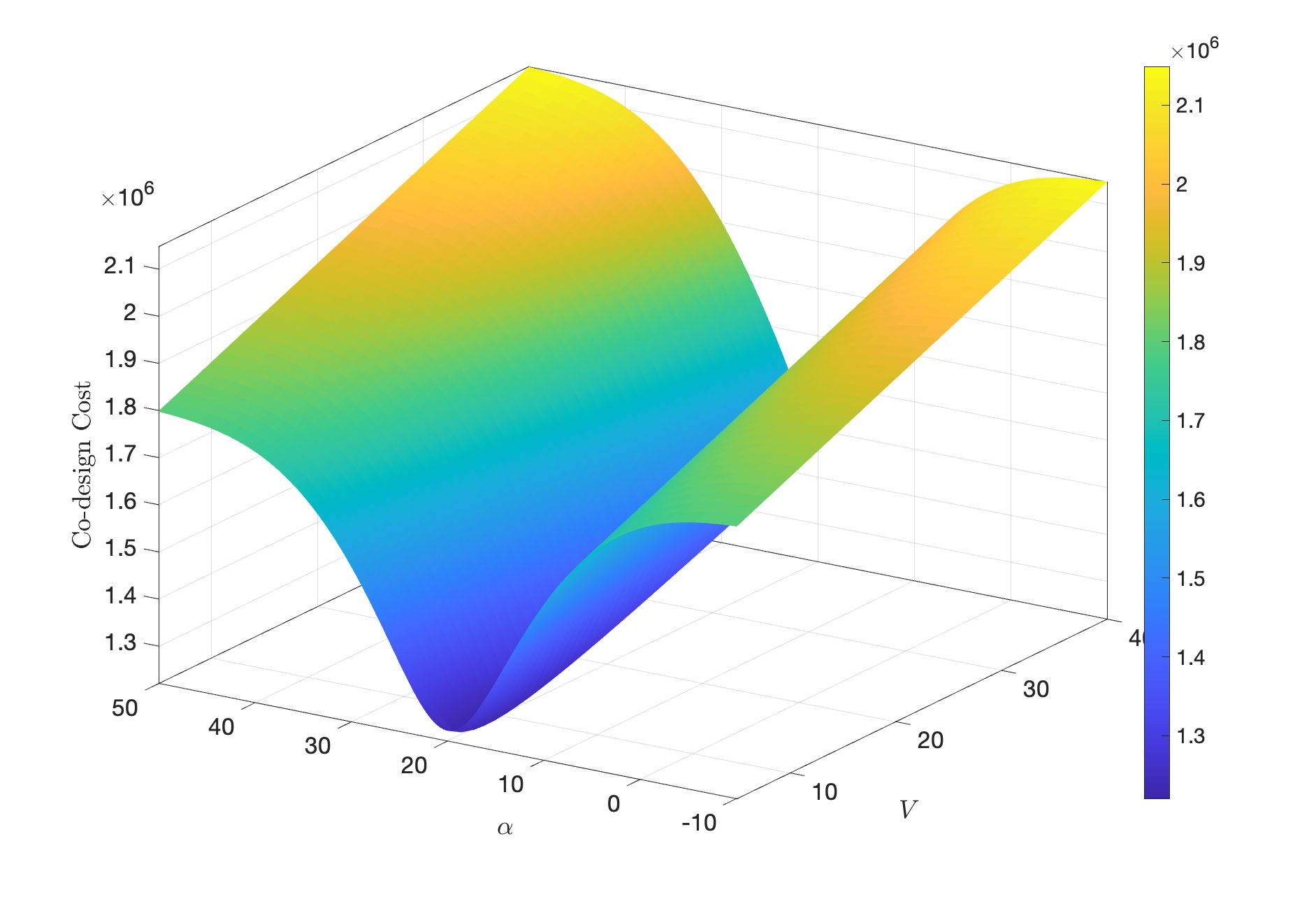}
    \caption{The co-design cost surface for a state-independent price threshold and constant demand.}
    \label{fig:codesign R1}
\end{figure}

Using the expression for $\pi^0(\alpha,V)$ in \eqref{eq:pi0 case 1}, the co-design optimization problem \eqref{prob:stochastic codesign 1} can easily be solved. The co-design cost surface is shown in Fig. \ref{fig:codesign R1}. It can be seen that the optimization problem is non-convex. However, for this example, the co-design cost does exist a unique minimum, allowing for targeted numerical optimization routines to be deployed.

The optimized parameters and costs are reported in Table~\ref{table:optimal solutions}. {In this example, since $\zeta=2$ and the demand is constant equal to 1, one needs in the long run to pump 50\% of the time, and not surprisingly the optimal state-independent $\alpha^*$ is equal to the mean $\mu=20$, which gives a pumping probability of $p=0.5$.} Next, we will compare these results to the case where the price thresholds are state-dependent.

\begin{table*}[!th]
    \centering
    \caption{Co-design solutions of three examples.}
    \begin{tabular}{lccc}    \toprule
     & State-independent price threshold & State-dependent price thresholds & State-dependent price thresholds\\
     & Constant demand & Constant demand & Uncertain demands \\
    \midrule
    Optimal tank size $V^*$ & 8 & 8 & 9.6 \\
    Optimal price threshold $\alpha^*$ & 20 & see Fig. \ref{fig:alpha ex2} & see Fig. \ref{fig:alpha ex3}\\
    Optimal capital cost & 80,000 & 80,000 & 96,000 \\
    Optimal operating cost over $N$ & 1,140,421  & 1,105,603 & 1,105,112 \\
    Optimal co-design cost $N$ & 1,220,421  & 1,185,603 & 1,201,112\\
    \bottomrule
\end{tabular}
\label{table:optimal solutions}
\end{table*}

%%%%%%%%%%%%%%%%%%%%%%%%%%%%%%%%%%%%%%%%%%%%%%%%%%%%%%%
\subsection{State-dependent Price Thresholds and Constant Demand}

\begin{figure}[th]
	\centering
	\includegraphics[width=\hsize]{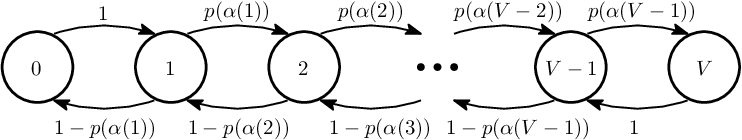}
	\caption{Markov chain for the example with state-dependent price thresholds and constant demand.}
	\label{fig:Example 2 markov chain}
\end{figure}

Here, the setting is the same as before, apart from that the electricity price thresholds are state-dependent. The Markov chain with transition probabilities is shown in Fig. \ref{fig:Example 2 markov chain}.

The Markov chain in Fig.~\ref{fig:Example 2 markov chain} is irreducible if all the transition probabilities $0< p(\alpha(i)) <1$, $i = 1,\ldots, V-1$. The stationary probabilities for each state in the Markov chain can be obtained by solving the normalization and balance equations.

As in the previous example (with $n_r = 0$), we follow the steps in Section \ref{section:codesign formulation} and obtain the following co-design optimization problem:
\begin{align}\label{prob:stochastic codesign 2}
    (\bm{\alpha}^*, V^*)  & :=  \underset{\bm{\alpha}, V} {\arg \min} \;\; {\frac{1}{N} c_v V} \nonumber\\
    &\; + \pi^{0}(\bm{\alpha},V) (\varepsilon_p\mu + w)  \nonumber\\
    &\; + \sum_{i = 1}^{V-1} \pi^{i}(\bm{\alpha},V) p(\alpha(i\Delta x)) \varepsilon_p \bar{c}(\alpha(i\Delta x)),
\end{align}
where $p(\alpha(i\Delta x))  = F(\alpha(i\Delta x))$, $\Delta x = d \Delta t =1$ and $\bm{\alpha} = [\alpha(i\Delta x), i = 1,\ldots, V-1]$.

\begin{figure}[t]
    \centering
    \subfigure[With constant demand]{\includegraphics[width=\hsize]{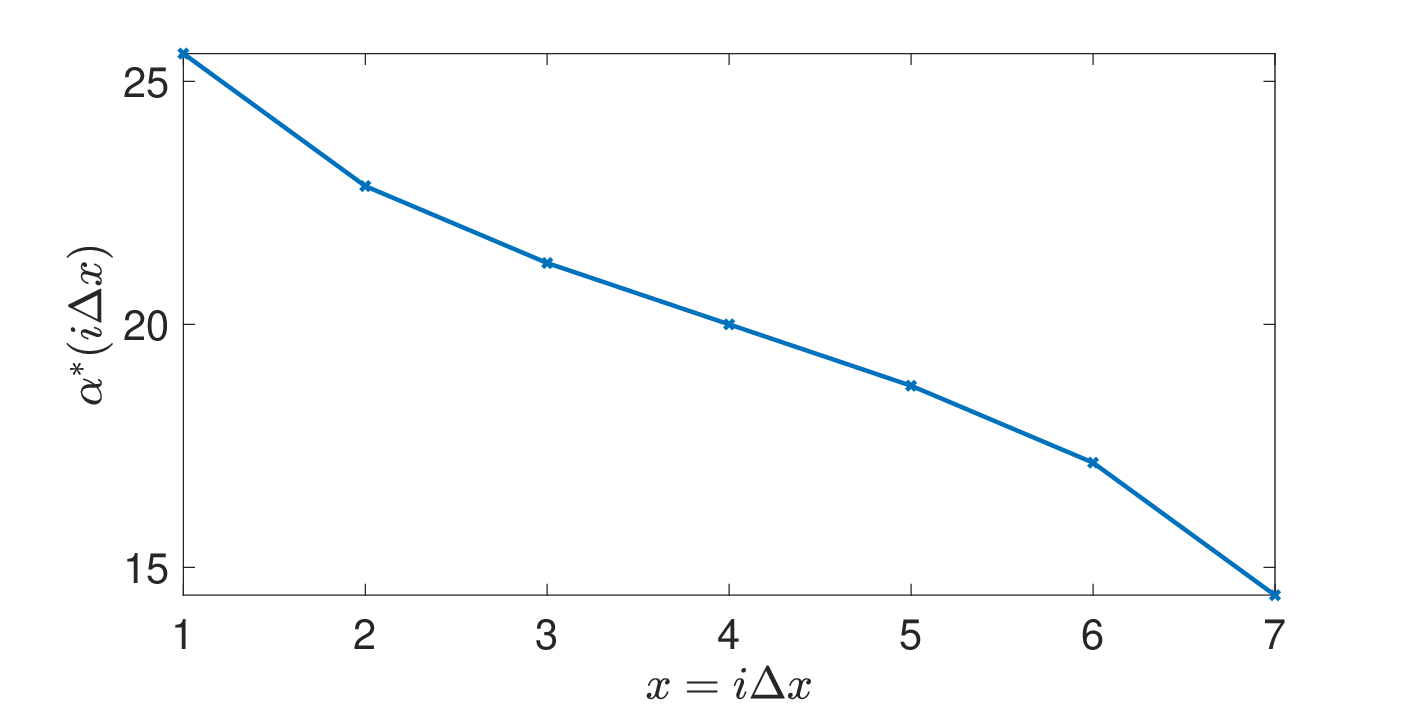}\label{fig:alpha ex2}}
    \subfigure[With uncertain demands]{\includegraphics[width=\hsize]{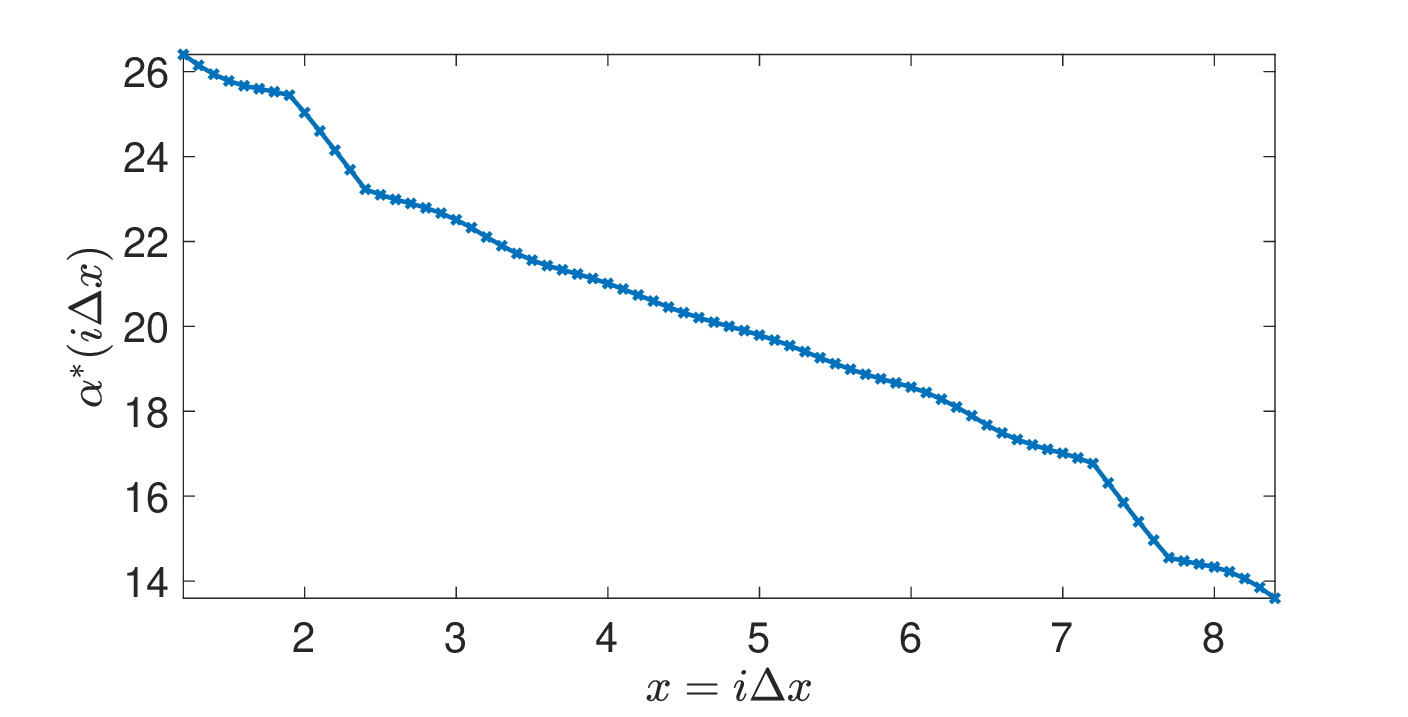}\label{fig:alpha ex3}}
    \caption{The optimal state-dependent price thresholds $\alpha^*(i \Delta x)$.}
    \label{fig:optimal state-dependent alpha}
\end{figure}

As the number of optimization parameters has increased due to state-dependent $\alpha(i \Delta x)$, the co-design surface can no longer be plotted, but a numerical solution can still be found. To obtain the solution to the co-design optimization problem~\eqref{prob:stochastic codesign 2}, we utilize Algorithm~\ref{algorithm:co-design} with the SPSA method proposed in \cite{spall2005introduction}. {The tank size $V$ varies from 5 to 30.} The optimal solutions and costs are presented in Table~\ref{table:optimal solutions} and the optimal state-dependent price thresholds are shown in Fig. \ref{fig:alpha ex2}. The observed trend reveals that as the volume in the tank increases, the threshold decreases. This trend is logical because when the volume in the tank is low, having larger threshold results in higher transition probabilities to higher volume states, consequently reducing the probabilities of transitioning to an even lower volume, which could potentially incur a significant operating cost due to enforced pumping (and additional penalties for running empty, although in this example $w=0$, so no explicit penalty is applied for running empty or below a given threshold volume). Similarly, when the volume is high there is less need to pump, and subsequently, a smaller price threshold can be set. 

Compared to the result in the previous example, the infrastructure cost is the same, but a lower co-design cost is achievable through the additional degrees of freedom available in the operating strategy.

\subsection{State-dependent Price Thresholds and Uncertain Demands}\label{subsection:state dependent alpha and uncertain demands}

To consider a more realistic example, the water demands are now uncertain:
\begin{align*}
    d_k = \begin{cases}
    0.8 & \text{with probability of } 0.2,\\
    0.9 & \text{with probability of } 0.2,\\
    1 & \text{with probability of } 0.2,\\
    1.1 & \text{with probability of } 0.2,\\
    1.2 & \text{with probability of } 0.2,\\
    \end{cases}
\end{align*}
for $k =1,\ldots, N$. Note the average demand is the same as in the previous examples. The demand quantization interval is chosen as $d = 0.1$ and the state quantization interval is $\Delta x = 0.1$.

Following the procedures in the previous example, the co-design optimization is formulated as follows:
\begin{align}\label{prob:stochastic codesign 3}
    (\bm{\alpha}^*, V^*) & :=  \underset{\bm{\alpha}, V} {\arg \min} \;\; {\frac{1}{N} c_v V} \nonumber\\
    &  +  \pi^0(\bm{\alpha},V) (\varepsilon_p\mu+w) + \sum_{j=1}^{n_{p}} \pi^{j}(\bm{\alpha},V) \varepsilon_p\mu \nonumber\\
    &  + \sum_{i = n_{p}+1}^{n_{s}} \pi^{i}(\bm{\alpha},V) p(\alpha(i\Delta x))  \varepsilon_p \bar{c}(\alpha(i\Delta x)),
\end{align}
where $p(\alpha(i\Delta x)) = F (\alpha(i \Delta x) )$,  $n_p = \frac{\underline{x}}{\Delta x} $ and $n_s = \frac{\overline{x}}{\Delta x} $ with $\underline{x} = 1.2 $ and $\overline{x} = V-1.2 $. $\bm{\alpha} = [\alpha(i\Delta x), i = n_p+1,\ldots, n_s]$.

We again use Algorithm \ref{algorithm:co-design} with the SPSA method to find the optimal co-design parameters reported in Table~\ref{table:optimal solutions} and Fig. \ref{fig:alpha ex3}. It is interesting to note that the state-dependent price threshold trend is very similar to the constant demand example, but the uncertain demand induces a more conservative infrastructure solution. Due to the larger tank, the operating cost is actually reduced compared to when the demand was constant.

\subsection{Sensitivity Analysis}\label{section:sensivitity all}

{As studied in \cite{Wang-2022}, Theorem \ref{theorem:long-run property} is demonstrated by the results from a Monte Carlo simulation consisting of 100 individual runs. The empirical operating costs are within 1\% of the expected operating cost in Theorem \ref{theorem:long-run property}.} In this section, we investigate the sensitivity of the results in the example in Section \ref{subsection:state dependent alpha and uncertain demands} with respect to variations in the distribution of energy prices. Two cases are considered:

In the first case, the sensitivity of the operating cost obtained from the optimal co-design solution with respect to the price distribution is investigated. The system is co-designed using constant assumed parameters $\bar{\mu}$ and $\bar{\sigma}$, but the true parameters $\mu$ and $\sigma$ are different from the assumed values. 

In the second case, the sensitivity of the achievable cost with the true parameters with respect to the assumed price distribution used in the co-design is investigated. The co-design is carried out for different parameters $\bar{\mu}$ and $\bar{\sigma}$ while the actual values $\mu$ and $\sigma$ are always kept constant.

\subsubsection{Sensitivity of Operating Cost to Changes in Actual Electricity Prices}
\label{sec:sens1}

Using Algorithm \ref{algorithm:co-design} with price distribution $\mathcal{N}(\bar{\mu}, \bar{\sigma})$, where $\bar{\mu} = 20$ and $\bar{\sigma} = 10$, the optimal co-design solutions are reported in Table \ref{table:optimal solutions} and Fig. \ref{fig:optimal state-dependent alpha}. {The tank size is fixed at the obtained optimal value $V^* = 9.6$. Using the obtained price thresholds, the operating cost can be directly computed using the result in Theorem \ref{theorem:long-run property} with different~$\mu $ and~$\sigma$.} The results are shown in Table \ref{table:sensitivity analysis 1}. 

\begin{table}[htbp]
    \centering
    \begin{threeparttable}
    \caption{Sensitivity results of operating cost to changes in actual electricity prices for fixed tank volume of 9.6.}
    \label{table:sensitivity analysis 1}
    \begin{tabular}{cccc}
        \toprule
        $\mu$ & $\sigma$ & {Expected Oper. Cost} & {Oper. Cost Diff.}  \\
        \midrule
        20 & 10 & {1,105,112} & 0.00\%   \\
        20 & 20 & {472,229}   & {-57.27\%} \\
        20 & 5  & {1,435,291} & {29.88\%}  \\
        24 & 10 & {1,503,717} & {36.07\%}  \\
        24 & 20 & {869,741}   & {-21.30\%} \\
        24 & 5  & {1,818,647} & {64.57\%}  \\
        16 & 10 & {802,917}   & {-27.35\%} \\
        16 & 20 & {168,941}   & {-84.71\%} \\
        16 & 5  & {1,117,847} & {1.15\% }  \\
        \bottomrule
    \end{tabular}
    \begin{tablenotes}
      \item The difference in operating cost is computed based on the cost with $\mu = 20$ and $\sigma = 10$ in the first row.
    \end{tablenotes}
    \end{threeparttable}
\end{table}

\begin{table*}[htbp]
    \centering
    \begin{threeparttable}
    \caption{Sensitivity results of total co-design cost to changes in electricity price distribution.}
    \label{table:sensitivity analysis 2}
    \begin{tabular}{cccccccc}
    \toprule
    $\tilde{\mu}$ & $\tilde{\sigma}$ & $V^*$ & Capital Cost & {Expected Oper. Cost} & {Total Cost} &  {Cost Diff.} \\
    \midrule
    20 & 10 & 9.60  & 96,000  & {1,105,112} & {1,201,112} & 0.00\%      \\
    20 & 20 & 12.30 & 123,000 & {1,095,060} & {1,218,060} & {1.41\%} \\
    20 & 5  & 7.50  & 75,000  & {1,142,901} & {1,217,901} & {1.40\%} \\
    24 & 10 & 9.60  & 96,000  & {1,153,317} & {1,249,317} & {4.01\%} \\
    24 & 20 & 12.30 & 123,000 & {1,111,419} & {1,234,419} & {2.77\%} \\
    24 & 5  & 7.50  & 75,000  & {1,214,616} & {1,289,616} & {7.37\%} \\
    16 & 10 & 9.60  & 96,000  & {1,153,317} & {1,249,317} & {4.01\%} \\
    16 & 20 & 12.30 & 123,000 & {1,111,424} & {1,234,424} & {2.77\%} \\
    16 & 5  & 7.50  & 75,000  & {1,214,616} & {1,289,616} & {7.37\%} \\
    \bottomrule
    \end{tabular}
    \begin{tablenotes}
      \item Total cost difference is computed based on the sum of capital cost and {expected operating cost from Theorem \ref{theorem:long-run property}} with $\tilde{\mu} = 20$ and $\tilde{\sigma} = 10$.
    \end{tablenotes}
    \end{threeparttable}
\end{table*}

Table \ref{table:sensitivity analysis 1} shows that if the mean price $\mu$ increases (or decreases) by 20\% from the mean used in design ($\tilde\mu=20$), the operating cost increases (or decreases) by a greater amount (around 36\% or -27\% respectively). Similarly, we note that the variance of the electricity price has a significant impact on the operating cost if there is a large deviation from the value used during the design. In summary, we can conclude that the actual operating cost can increase or decrease significantly if the actual price distribution is different from the one used during the design.

\subsubsection{Sensitivity of Co-design Optimization to Changes in Electricity Price Distributions}\label{sec:sensitivity2}

Here we investigate the sensitivity of the co-design method by fixing the price distribution encountered at $\mu=20$ and $\sigma=10$, and using incorrect parameters $\tilde{\mu}$ and $\tilde{\sigma}$ during the design. The results are reported in Table \ref{table:sensitivity analysis 2}.

\begin{figure}[t]
    \centering
    \includegraphics[width=\hsize]{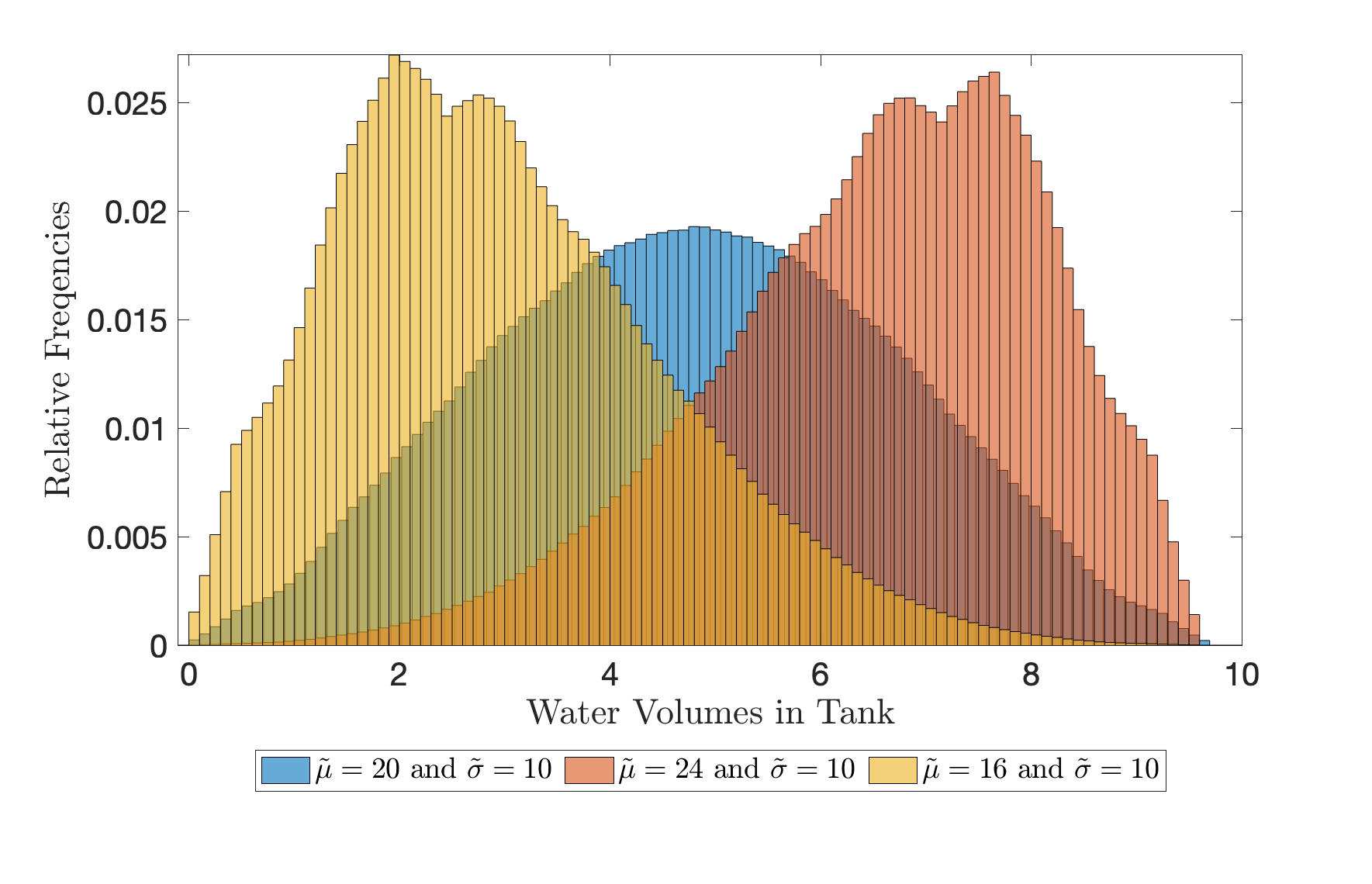}
    \caption{{Comparison of stationary probabilities} with different $\tilde{\mu}$.}
    \label{fig:relative frequencies compare 1}
\end{figure}

For cases with the same $\tilde{\sigma} = 10$ but increased $\tilde{\mu}$, fewer enforced pumpings are triggered and as expected more time is spent in high tank volume states, as observed in Fig.~\ref{fig:relative frequencies compare 1}. The converse is also true when $\tilde{\mu}$ is decreased. Nonetheless, the overall cost difference is relatively low, and the capital cost is the same in each case.

As can be seen from Table \ref{table:sensitivity analysis 2}, the total actual cost is rather insensitive to the price distribution used during the design phase, that is the obtained $V$ and $\alpha(i\Delta x)$ also work well when the actual price distribution is different. The largest difference occurs when the standard deviation~$\tilde{\sigma}$ is underestimated, and this is due to that the co-design optimization selects a smaller tank size which leads to more frequent enforced pumping events. 

To conclude, the first case investigated in  Section \ref{sec:sens1} shows that the operating cost achieved in the design phase is sensitive with respect to variations in the assumed price distribution. The results in this section show that, even if we had known the true distribution, we would not have been able to improve significantly on the actual cost.

%%%%%%%%%%%%%%%%%%%%%%%%%%%%%%%%%%%%%%%%%%%%%%%%%%%%%%%
%%%%%%%%%%%%%%%%%%%%%%%%%%%%%%%%%%%%%%%%%%%%%%%%%%%%%%%
\section{Case Study: a Water Network in South Australia}\label{section:case study}

In this section, we apply the proposed co-design method to a high-fidelity simulation of a real-world water network in South Australia, operated by the South Australian Water (SA Water). We first describe the system and then present the data processing procedure for obtaining the parameters required for solving the co-design optimization problem. Then, we describe the simulation setup, which makes use of an EPANET hydraulic model. Finally, we evaluate the effectiveness of the solutions through a comparison with historic operational data from 2019. 

\subsection{Description}

The network topology is aggregated to be consistent with that shown in Fig. \ref{fig:example}, so that it includes a pumping station with one pump, one storage tank, and an aggregated demand sector representing the combined demand of all downstream sectors. The co-design objective is to determine the optimal combined tank size and the price thresholds for operating the pumping station using the control policy in \eqref{eq:pumping flow}.

\begin{figure}[t]
    \centering    
    \includegraphics[width=\hsize]{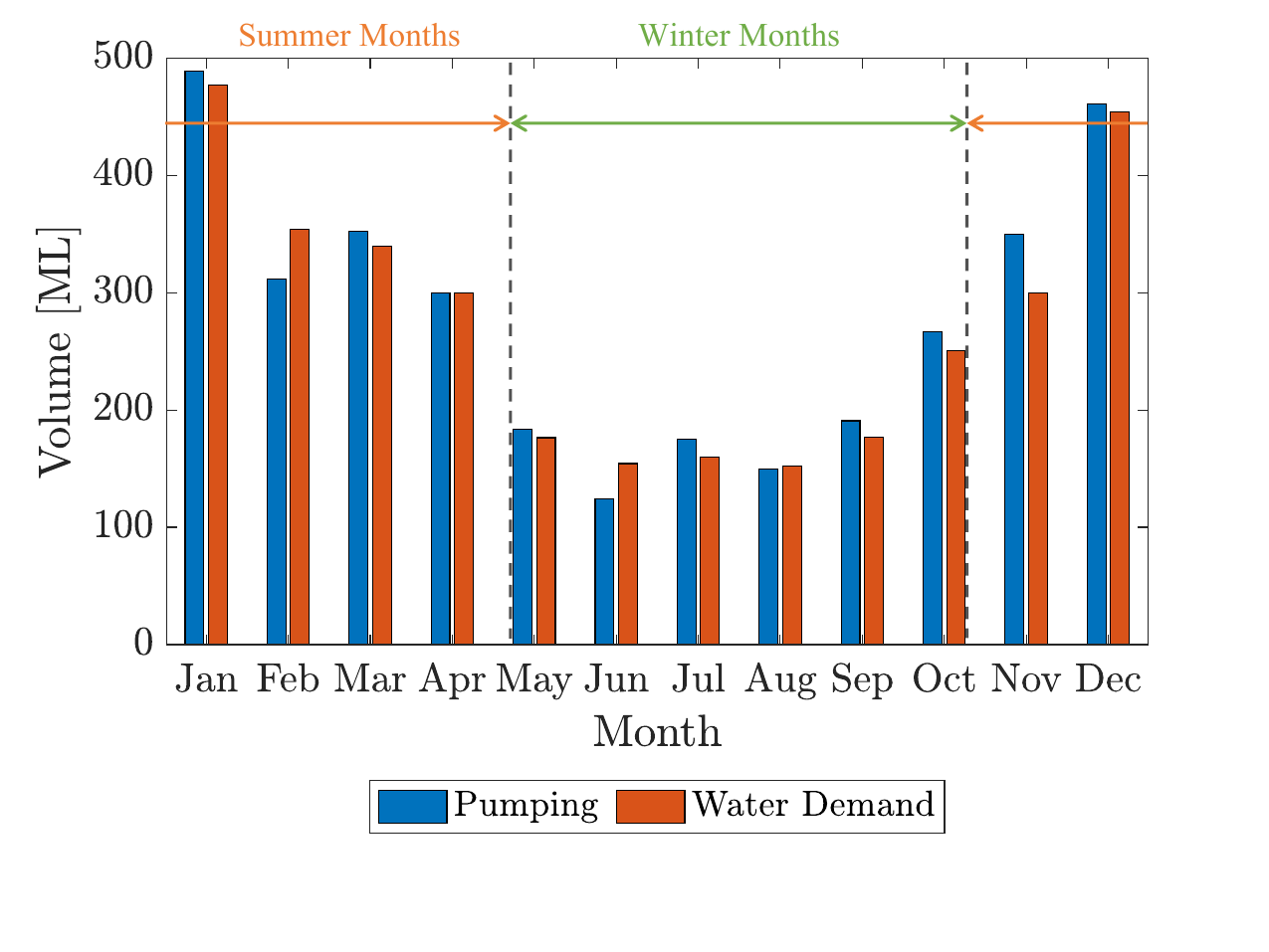}
    \caption{Monthly pumping and water demand from historical data with operation by SA Water using a trigger-level control.}
    \label{fig:monthly data}
\end{figure}

For this water network, pumping flows and water demands in 2019 are available. The SA electricity prices in 2019 are available from the AEMO with a sampling time of 30 minutes~\cite{nem_data}. As shown in Fig. \ref{fig:monthly data}, the water demand increases in the warmer months since the network services popular holiday area. The year was therefore divided into a summer period from November to April and a winter period from May to October.

\begin{figure}[ht]
    \centering
    \includegraphics[width=0.9\hsize]{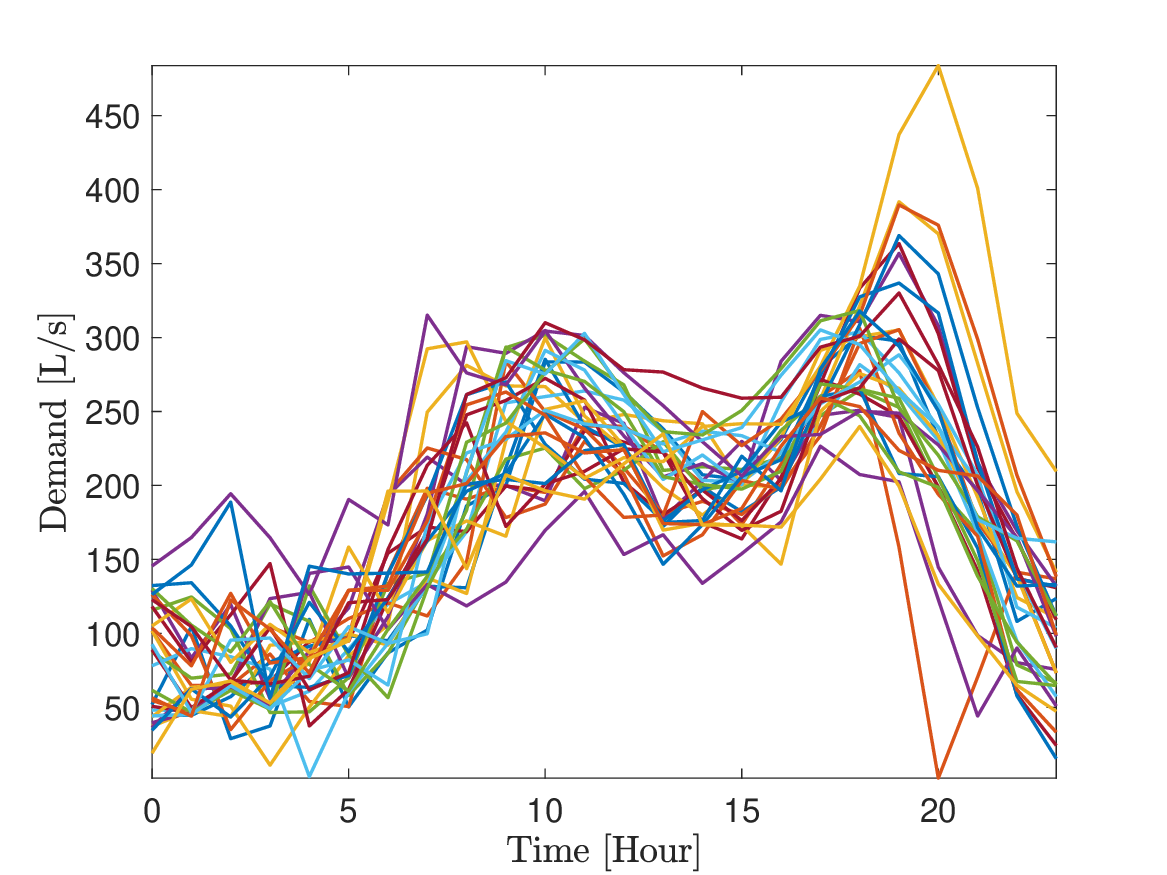}
    \caption{Typical summer-month daily water demands in January 2019 from historical data. {Each curve represents the water demand for a particular day from midnight.}}
    \label{fig:demand}
\end{figure}

\begin{figure}[thpb]
    \centering
    \includegraphics[width=0.9\hsize]{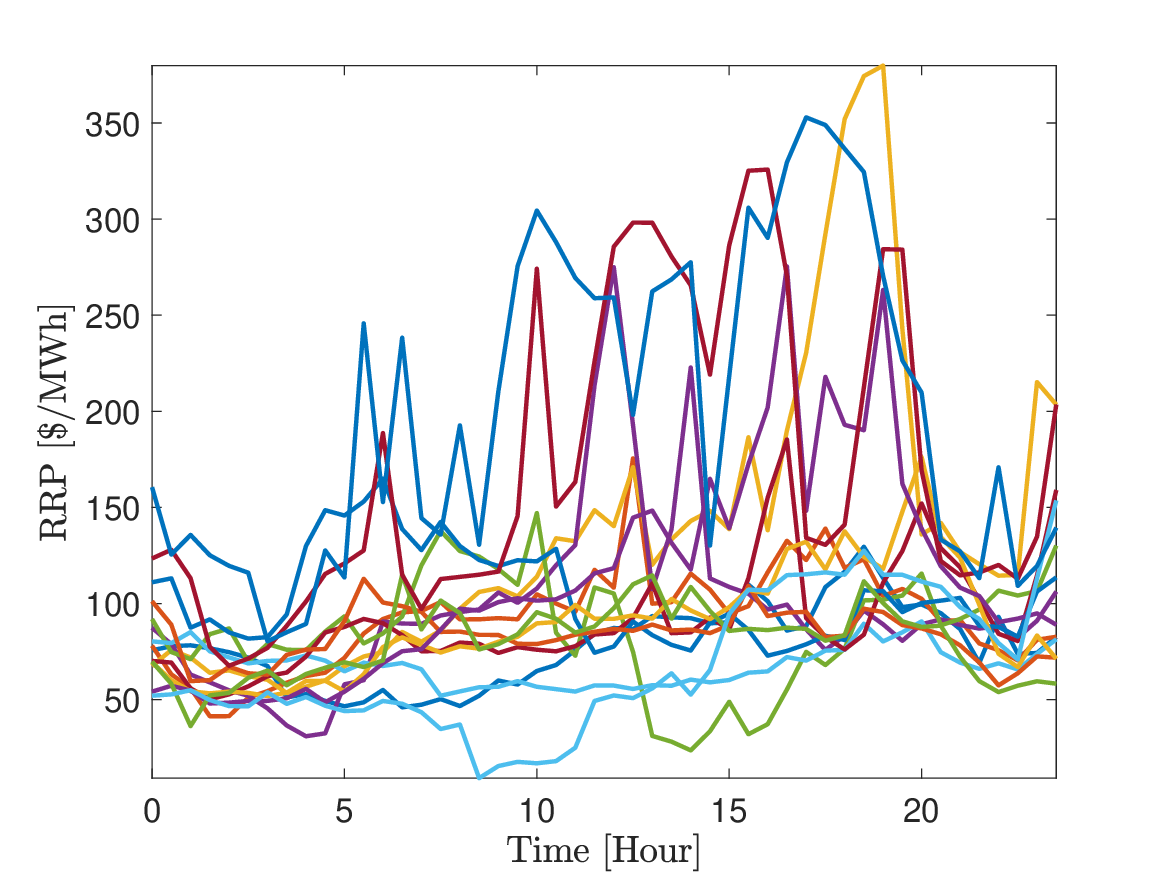}
    \caption{Typical South Australian electricity retail prices during January 2019 sourced from \cite{AEMO}.}
    \label{fig:electricity price}
\end{figure}

The following parameters were set based on available data:
\begin{itemize}
    \item {The sampling time is $\Delta t = 1 $ hour.}
    \item Water demand has a noticeable daily pattern shown in Fig.~\ref{fig:demand}. The period $T=24$ hours is therefore used. 
    \item The quantization interval is chosen as $ d = 43$ L/s, and the quantized water demands for summer and winter months are chosen as $\tau d$ with $\tau = 1,\ldots, 11$. The corresponding probabilities $a_{\kappa}^{\tau}$ of demands were estimated for every $\kappa = \mod (k,T) = 0, \ldots, 23$.
    \item When the pump operates, the flow is $q = 215$~L/s. {This indicates $\zeta = 5$ in \eqref{eq:q def}.}
    \item SA Water purchase electricity directly from the electricity market, with prices available in the AEMO database~\cite{AEMO}. Some examples of how electricity prices vary over a 24-hour period are shown in Fig.~\ref{fig:electricity price}. Extreme price events, which are taken to be when prices are above \$500 per MWh, are removed for investigating price distributions. While actual price histories are used in the simulation, for the design they are assumed to be independently and identically distributed Gaussian random variables. Two distributions can be estimated by using data in the summer and winter months, respectively. For summer months, the mean electricity price is $\mu = \$89.77$/MWh and the standard deviation is $\sigma = \$43.39 $/MWh while for winter months, $\mu = \$78.57$/MWh and $\sigma = \$42.58 $/MWh.
    \item Water storage tanks with different sizes are considered in the co-design problem. Overall, the life cycle of the tank is taken to be 50 years \cite{WSAA2011}. The options for tank sizes are 3, 4, 5, 8, 10, 15 and 20 ML. The corresponding capital costs can be found in Table~\ref{table:co-design result} based on the numbers reported in \cite{NSW2014}. 
    \item {For a tank size $V$, $n = \frac{V}{\Delta x}$, where $\Delta x = d \Delta t = 0.1548$~ML. Furthermore, $n_p = \max (\tau) - 1$ and $n_s = n - \max(\tau)$.}
    \item The penalty cost is \$10,000 per times when the tank is empty.
\end{itemize}

\subsection{{Simulation Results}}

\begin{figure}[t]
	\centering
        \includegraphics[width=\hsize]{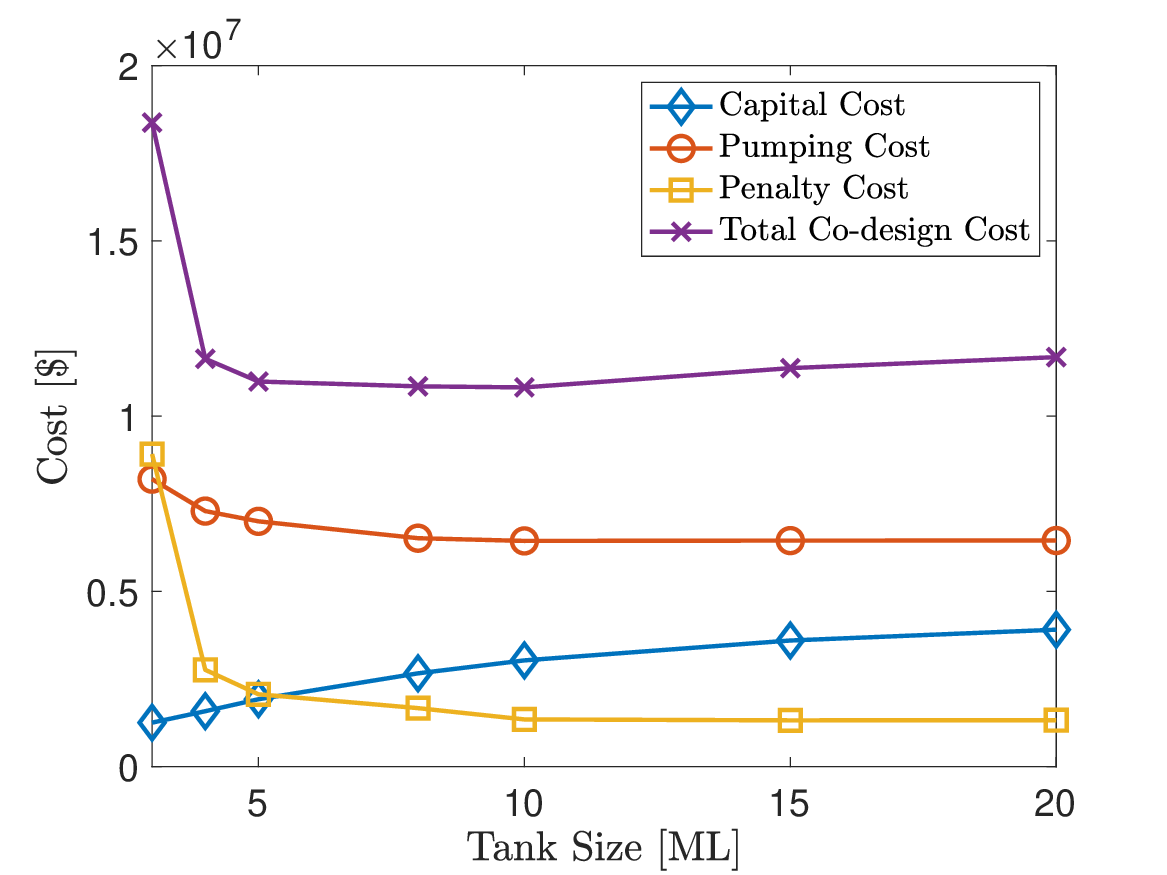}
	\caption{The cost comparison for a 50-year planning horizon with different tank sizes and optimal co-design control operations.}
	\label{fig:codesign cost-tank size}
\end{figure}

As the infrastructure lifetime is set to 50 years, the operating cost is found by using the summer and winter parameters for 25 years each. For each tank size, the total cost is found by adding the capital cost and the operating cost. The results are reported in Table \ref{table:co-design result}. As also shown in Fig. \ref{fig:codesign cost-tank size}, it can be seen that a smaller tank size may save on capital costs but leads to higher operating and penalty costs. When the tank is too small, the risk of having less water in the tank than the minimum allowed increases. A larger tank provides more flexibility in storing water and meeting demands during high-priced times, but the savings in operating costs may not compensate for the increase in capital costs. 

When the tank sizes are greater than 8 ML, the operating costs are similar since further increases in tank size offer no further improvement on operating cost under the considered control policy. In general, as shown in Fig. \ref{fig:codesign cost-tank size}, the co-design optimization provides a balanced solution for tank size and control parameters. The optimal solution for the tank size is $V^* = 10$ ML with a minimum total co-design cost over the planning horizon of 50 years. From Fig. \ref{fig:codesign cost-tank size}, it can be seen that the total costs are quite flat for large tank sizes. Therefore, in practice one could consider a larger tank size, which provides some insurance against larger variations in electricity prices as discussed in Section \ref{section:sensivitity all}.

\begin{table*}[!ht]
    \centering
    \caption{Co-design computation results for the case study network.}
    \begin{tabular}{cccccc}
    \toprule
    Tank Size   {[}ML{]} & {Number of States} & Capital Cost & Pumping Cost & Penalty Cost & Total Co-design Cost \\
    \midrule
    3                    & {480}                           & \$1,256,052  & \$8,202,120          & \$8,916,983        & \$18,375,154         \\
    4                    & {624}                            & \$1,582,082  & \$7,291,233          & \$2,763,045        & \$11,636,360         \\
    5                    & {792}                          & \$1,923,676  & \$6,999,010          & \$2,065,091        & \$10,987,777         \\
    8                    & {1248}                           & \$2,662,437  & \$6,517,730          & \$1,669,709        & \$10,849,875         \\
    10                   & {1560}                          & \$3,031,888  & \$6,440,382          & \$1,347,709        & \$10,819,979         \\
    15                   & {2328}                         & \$3,599,156  & \$6,449,178          & \$1,321,563        & \$11,369,897         \\
    20                   & {3120}                         & \$3,908,106  & \$6,451,298          & \$1,324,352        & \$11,683,756  \\
    \bottomrule
    \end{tabular}
    \label{table:co-design result}
\end{table*}

\begin{table*}[!ht]
    \centering
    \begin{threeparttable}
    \caption{Comparison of historical and co-design operation by simulation for the year of 2019.}
    \label{table:comparison SA Water}
    \begin{tabular}{lcccccc}
    \toprule
    \multicolumn{1}{l}{} & \multicolumn{2}{c}{Historical Operation} & \multicolumn{2}{c}{Co-design Operation} &              & \multicolumn{1}{l}{} \\ \cline{2-5}
    & Energy Cons. {[}kWh{]}  & Energy Cost  & Energy Cons. {[}kWh{]}   & Energy Cost  & Cost Saving & Percentage     \\ \midrule
    Summer   Months & 1,305,914 & \$142,083 & 1,339,054 & \$123,436 & -\$18,647 & -13\% \\
    Winter Months   & 622,756   & \$46,977  & 662,965   & \$31,102  & -\$15,875 & -34\% \\
    Whole year      & 1,928,670 & \$189,060 & 2,002,019 & \$154,538 & -\$34,522 & -18\%  
    \\ \bottomrule
    \end{tabular}
    \begin{tablenotes}
      \item Cost Savings = Energy Cost (Co-design Operation) - Energy Cost (Historical Operation), 
      \item Percentage  = Cost Saving/Energy Cost (Historical Operation).
    \end{tablenotes}
    \end{threeparttable}
\end{table*}

\subsection{Comparison of Operations with Existing Infrastructure}

The proposed co-design method can also be used to improve control operations for existing infrastructure, leveraging the optimal control parameters for a given tank size. Using historical data on water demands and electricity prices in 2019, we compared the co-design solutions and historical operations in 2019. The historical operation was based on trigger-level control. It should also be noted that our optimization considers 2019 data only, while SA Water strategy may have considered different metrics and risk scenarios over a longer time. Nevertheless, the comparison is believed to serve as a reasonable representation of potential improvements to current practice. For this case study, the tank size is 38.5 ML, in which 28.5 ML is an emergency buffer. Moreover, the maximum volume of the existing storage is 136 ML. An EPANET hydraulic model of the case study was used in the closed-loop simulation. Table \ref{table:comparison SA Water} shows the results. 

\begin{figure}[t]
	\centering
	\includegraphics[width=\hsize]{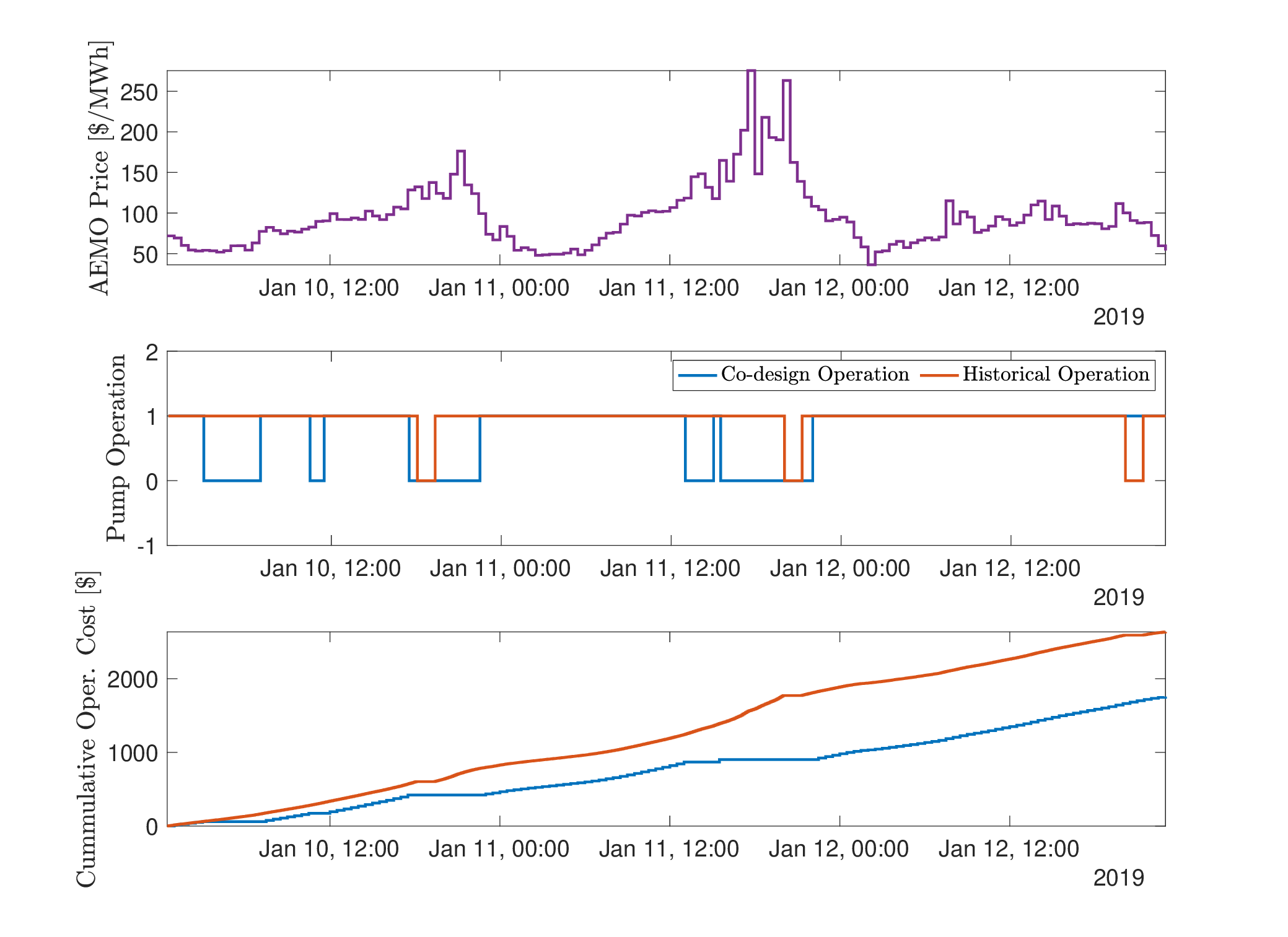}
	\caption{Comparison of historical and co-design operation based on January 2019 data obtained from SA Water network.}
\end{figure}

During the summer period in 2019, the operation using the optimized price thresholds for the given tank size (referred as co-design operation) resulted in a 13\% decrease in pumping cost relative to trigger-level operations, while a 34\% decrease is observed during the winter months. {The reason for the larger savings in the winter months is that there are more opportunities to shift pumping from high price periods to low price periods when the demand is low.} Overall, the co-design solutions saved 18\% in pumping costs for the year 2019. The considered control policy based on the price threshold is hence effective and able to bring economic benefit to the operation of the existing water infrastructure.

%%%%%%%%%%%%%%%%%%%%%%%%%%%%%%%%%%%%%%%%%%%%%%%%%%%%%%%%%%%%%%%%%%%%%%%
%%%%%%%%%%%%%%%%%%%%%%%%%%%%%%%%%%%%%%%%%%%%%%%%%%%%%%%%%%%%%%%%%%%%%%%
\section{Conclusions}\label{section:conclusions}

In this paper, we have proposed a tractable stochastic co-design optimization method that simultaneously determines the optimal tank storage size and control parameters used in the considered control policy. {The co-design optimization leverages asymptotic Markov chain theory and employs some conditions regarding the stochastic nature of electricity prices and water demands. Through three examples and a real-world case study, the effectiveness of the proposed co-design method has been demonstrated. Notably, the case study shows that the co-design method also brings economic benefits to the operation of existing water infrastructure by finding the optimal price thresholds given the existing infrastructure.} As future research, the assumptions about the electricity prices and demands will be relaxed allowing for time dependencies that more accurately reflect the stochastic nature of actual electricity prices and water demands.

% or


% \appendices
% \section{Markov Chain for the Co-design Problem in Section III}


% % you can choose not to have a title for an appendix
% % if you want by leaving the argument blank
% \section{}

\section*{Acknowledgment}

Dr Ye Wang is supported by the Australian Research Council through the 2022 Discovery Early Career Researcher Award (DE220100609). The authors also thank SA Water for providing the hydraulic model and data for the case study.

%We also acknowledge Dr~Lisa Blinco from SA Water for providing case study data and useful discussions during this research project.

\ifCLASSOPTIONcaptionsoff
  \newpage
\fi



\bibliographystyle{IEEEtran}
\bibliography{IEEEabrv,codesign}



% \end{thebibliography}


% \end{IEEEbiography}


%\newpage



\vfill


\end{document}